\pdfoutput=1
\documentclass[
  prl,
  twocolumn,
  superscriptaddress,
  longbibliography,
  nofootinbib,
  nobibnotes
]{revtex4-2}

\usepackage{amsmath,amssymb,bm}
\usepackage{xcolor}
\usepackage{graphicx,dcolumn,braket,slashed}
\usepackage[colorlinks=true,linkcolor=blue,citecolor=blue,urlcolor=blue]{hyperref}
\usepackage{bookmark}

\usepackage{mathtools}
\usepackage{multirow}
\usepackage{comment}
\usepackage{array}
\usepackage{textcomp}
\usepackage{dsfont}
\usepackage{longtable}
\usepackage{chngcntr}

\newcommand{\Ree}{\operatorname{Re}}

\newcommand{\avgw}[1]{\left\langle\!\left\langle #1 \right\rangle\!\right\rangle}

\newcommand{\Evec}{\bm{E}}                 
\newcommand{\Hvec}{\bm{H}}                 
\newcommand{\EpsMat}{\bm{\varepsilon}}     

\newcommand{\avgT}[1]{\langle #1 \rangle_T}

\newcommand{\be}{\begin{eqnarray}}
\newcommand{\ee}{\end{eqnarray}}
\newcommand{\bbm}{\begin{bmatrix}}
\newcommand{\ebm}{\end{bmatrix}}
\newcommand{\bpm}{\begin{pmatrix}}
\newcommand{\epm}{\end{pmatrix}}

\makeatletter
\g@addto@macro\normalsize{%
  \setlength\abovedisplayskip{12pt}%
  \setlength\belowdisplayskip{12pt}%
  \setlength\abovedisplayshortskip{4pt}%
  \setlength\belowdisplayshortskip{4pt}%
}
\makeatother

\usepackage[normalem]{ulem}

\begin{document}

\title{Energy Transport Velocity in Photonic Time Crystals}

\author{Kyungmin Lee}
\thanks{These authors contributed equally to this work.}
\affiliation{Department of Physics, Korea Advanced Institute of Science and Technology, Daejeon 34141, Republic of Korea}

\author{Younsung Kim}
\thanks{These authors contributed equally to this work.}
\affiliation{Department of Physics, Korea Advanced Institute of Science and Technology, Daejeon 34141, Republic of Korea}

\author{Kun Woo Kim}
\email{kunx@cau.ac.kr}
\affiliation{Department of Physics, Chung-Ang University, 06974 Seoul, Republic of Korea}

\author{Bumki Min}
\email{bmin@kaist.ac.kr}
\affiliation{Department of Physics, Korea Advanced Institute of Science and Technology, Daejeon 34141, Republic of Korea}

\date{\today}

\begin{abstract}
Steep or near-vertical Floquet dispersion in photonic time crystals (PTCs) is often read as fast, even apparently superluminal, transport. Here, we demonstrate that this anomaly arises from modulation-driven geometric drift, not energy flow. Using the electromagnetic power-flux pairing in the Maxwell–Floquet formulation, we prove that the cycle-averaged energy velocity remains strictly bounded. We further establish a universal velocity-product law conserved throughout the passband, $ v_E v_g=\langle v_{\rm ph}^2\rangle_T $, fixing transport solely by the temporal average of the inverse permittivity. The divergent group velocity is then traced to a mismatch between electric and magnetic geometric phase connections, revealing apparent superluminality as a geometric effect of temporal modulation.
\end{abstract}

\maketitle

\emph{Introduction.—}
Since the early analyses of Sommerfeld and Brillouin, it has been clear that in dispersive media the group velocity
\(v_g \equiv d\omega/dk\) of a pulse can exceed the vacuum speed of light \(c\), vanish, or even become negative without violating relativity~\cite{sommerfeld1914fortpflanzung,brillouin2013wave,wang2000gain}.
This motivates a hierarchy of velocities, including the energy-transport velocity and the signal (front) velocity~\cite{brillouin2013wave,oughstun2012electromagnetic,peatross2000average}.
In passive, causal media, the front velocity is identically equal to the vacuum speed of light $c$ due to the vanishing material response at the high-frequency limit~\cite{sommerfeld1914fortpflanzung,brillouin2013wave}, while the energy-transport velocity remains strictly bounded (and typically subluminal), whereas \(v_g\) can take anomalous values~\cite{milonni2004fast,boyd2009controlling,huang2001poynting,kuzmich2001signal}.
Modern coherent-control and dispersion-engineering platforms, ranging from electromagnetically induced transparency to photonic crystals, enable extreme group indices and widely tunable \(v_g\)~\cite{PhysRevLett.64.1107,hau1999light,baba2008slow,d2001group}.
Thus, slow- and fast-light experiments consistently show that \(d\omega/dk\) is a convenient but sometimes misleading proxy for transport, and must be distinguished from the velocities governing the flow of energy and information~\cite{garrett1970propagation,chu1982linear,stenner2003speed}.

Recently, time-modulated photonic media~\cite{caloz2019spacetime,park2021spatiotemporal,galiffi2022photonics,pacheco2022time,asgari2024theory,park2021spatiotemporal,horsley2023quantum,pendry2024qed,park2025spontaneous,bae2025quantum} have renewed interest in wave dynamics in driven settings. Photonic time crystals (PTCs)~\cite{PhysRevB.98.085142,park2021spatiotemporal,asgari2024theory,park2025spontaneous,sustaeta2025quantum,bae2025quantum,allard2026broadband}, spatially homogeneous media with \(\varepsilon(t+T)=\varepsilon(t)\), are described by a Maxwell--Floquet eigenproblem that is often cast as a non-Hermitian problem  in a Floquet-harmonic representation~\cite{shirley1965solution,sambe1973steady}. Their quasifrequency bands \(\omega(k)\) exhibit passbands and momentum gaps with complex \(\omega\), signaling parametric growth or decay~\cite{tien1958parametric,lee2021parametric,park2022revealing,wang2025expanding}; near momentum gap edges the modes are standing waves and \(d\omega/dk\) can diverge~\cite{martinez2017standing}, a feature recently confirmed in microwave PTC band-structure measurements~\cite{reyes2015observation,park2022revealing,wang2023metasurface,xiong2025observation,hou2026experimental,zhu2025spatiotemporal}. Related Floquet platforms also report near-vertical dispersion~\cite{4lqd-z567}, and other superluminal-looking dynamics~\cite{zou2024momentum,pan2023superluminal}. This raises a transport-level question: in a PTC, what velocity truly diagnoses energy transport, and can any superluminal-looking feature correspond to energy flow exceeding the relevant phase-velocity bound?

In this Letter, we work within Maxwell theory for a PTC with strictly positive, time-periodic permittivity. 
In the Floquet-harmonic representation, the Maxwell symplectic form, evaluated with the derivatives of the Maxwell–Floquet matrix, reduces to the cycle-averaged Poynting flux and energy density: this defines a symplectic velocity and proves $v_{\rm symp}=v_E$ on any simple real Floquet band, hence $|v_E|$ is bounded by the maximum instantaneous phase velocity.
A localized wavepacket can drift either by true Poynting-flux transport or by modulation-induced reweighting of the energy density. This separation resolves the near-vertical band slopes: $v_g=d\omega/dk$ may diverge at gap edges even when energy transport is suppressed. Indeed, on the entire passband we prove the exact constraint
$ v_E v_g=\langle(\mu\varepsilon(t))^{-1}\rangle_T=\langle v_{\rm ph}^2\rangle_T $, so the group-velocity divergence is precisely compensated by $v_E\to 0$. In $k$-space, the modulation-driven drift is governed by a gauge-invariant mismatch between the electric and magnetic phase connections of the Floquet eigenmodes.

\emph{Model and energy velocity.—}
We consider one-dimensional propagation along \(z\) in a spatially uniform, isotropic, source-free medium with constant \(\mu>0\) and real, \(T\)-periodic permittivity \(\varepsilon(t)\ge\varepsilon_{\min}>0\). Vacuum constants are denoted \(\varepsilon_v,\mu_v\), so \(c\equiv 1/\sqrt{\varepsilon_v\mu_v}\), and we assume \(\mu\,\varepsilon_{\min}\ge \mu_v\,\varepsilon_v\) (equivalently, the maximum instantaneous phase velocity does not exceed \(c\)). Working in complex notation, the instantaneous electromagnetic energy density and Poynting flux are
\(u(z,t)=\tfrac14\bigl(\varepsilon(t)|E(z,t)|^2+\mu|H(z,t)|^2\bigr)\) and \(S_z(z,t)=\tfrac12\Ree\!\{E(z,t)\,H^*(z,t)\}\), which satisfy the Poynting theorem
\(\partial_t u+\partial_z S_z=-\tfrac14\,\dot\varepsilon(t)\,|E(z,t)|^2\), expressing local energy exchange between the field and the modulation~\cite{pendry2021gain}.
We denote by \(\avgT{\cdot}\) the average over one modulation period and define the cycle-averaged energy-transport velocity 
\begin{equation}
v_E \equiv \frac{\avgT{S_z}}{\avgT{u}}.
\label{eq:vE-def}
\end{equation}
The pointwise inequality \(|S_z|\le u/\sqrt{\varepsilon\mu}\) implies \(|v_E|\le v_{\mathrm{ph,max}}\equiv 1/\sqrt{\mu\,\varepsilon_{\min}}\le c\),
so the energy-transport velocity remains finite even when the Floquet dispersion becomes extremely steep~\cite{SM}. Throughout this Letter we restrict to nondispersive media with real, strictly positive \(\varepsilon(t)\); extensions are discussed in the Supplemental Material~\cite{SM}.

\emph{Maxwell--Floquet matrix and symplectic velocity.—}
To connect the transport velocity to the Floquet band structure, we use the plane-wave ansatz
\(E(z,t)=E(t)e^{ikz}\), \(H(z,t)=H(t)e^{ikz}\) with real \(k\), which gives
\(ik\,E(t)=-\mu\,\partial_t H(t)\) and \(ik\,H(t)=-\partial_t\!\big(\varepsilon(t)E(t)\big)\).
With \(\varepsilon(t+T)=\varepsilon(t)\), we expand
\(E(t)=\sum_m E_m e^{-i\omega_m t}\), \(H(t)=\sum_m H_m e^{-i\omega_m t}\),
\(\varepsilon(t)=\sum_\ell \varepsilon_\ell e^{-i\ell\Omega t}\),
where \(\Omega=2\pi/T\) and \(\omega_m=\omega+m\Omega\).
Matching harmonics yields
\begin{equation}
kE_m=\mu\,\omega_m H_m,\qquad
kH_m=\omega_m\sum_n \varepsilon_{m-n}E_n.
\label{eq:harmonics}
\end{equation}
Collecting the coefficients into $|R\rangle=(\Evec,\Hvec)^T$, where $\Evec=\{E_m\}$ and $\Hvec=\{H_m\}$, gives the Maxwell--Floquet eigenvalue problem, written as the matrix equation
\begin{equation}
M(\omega,k)\,\ket{R} = 0.
\label{eq:MR=0}
\end{equation}
Nontrivial solutions exist only when $M(\omega,k)$ is singular; this singularity condition determines the Floquet dispersion $\omega(k)$. Real roots define passbands, whereas complex roots define momentum gaps.

On a simple band, we introduce the left null vector \(\bra{L}\) of the non-Hermitian matrix $M(\omega,k)$, defined by \(\bra{L}M(\omega,k)=0\). Differentiating \(M(\omega(k),k)\ket{R(k)}=0\) gives the exact slope formula~\cite{JohnsonIb02}
\begin{equation}
v_g(k)\equiv \frac{d\omega}{dk}
=
-\frac{\bra{L}\,\partial_k M\,\ket{R}}{\bra{L}\,\partial_\omega M\,\ket{R}}.
\label{eq:vg-geo}
\end{equation}
Equation~\eqref{eq:vg-geo} gives the geometrical slope of the Floquet dispersion and depends on the left null vector \(\bra{L}\) of the non-Hermitian matrix.

In Maxwell theory, however, the physically natural pairing is fixed by the symplectic form of the source-free Maxwell equations. With the symplectic matrix
\begin{equation}
\mathbb J = \begin{pmatrix}
0 & \mathbb I\\[2pt]
-\mathbb I & 0
\end{pmatrix},
\label{eq:J}
\end{equation}
which pairs the electric and magnetic harmonic sectors, we define the symplectic velocity
\begin{equation}
v_{\rm symp}(k)\equiv
-\frac{\bra{R}\,\mathbb J\,\partial_k M\,\ket{R}}
{\bra{R}\,\mathbb J\,\partial_\omega M\,\ket{R}}.\label{eq:vsymp}
\end{equation}
Although this has a Hellmann--Feynman--like structure, it is evaluated with \(\bra{R}\mathbb J\), rather than with the left null vector \(\bra{L}\). As shown below, this \(\mathbb J\)-weighted evaluation connects the Maxwell--Floquet matrix derivatives to the cycle-averaged flux and energy.

\begin{figure}[htb!]
  \centering
    \includegraphics[width=0.45\textwidth]{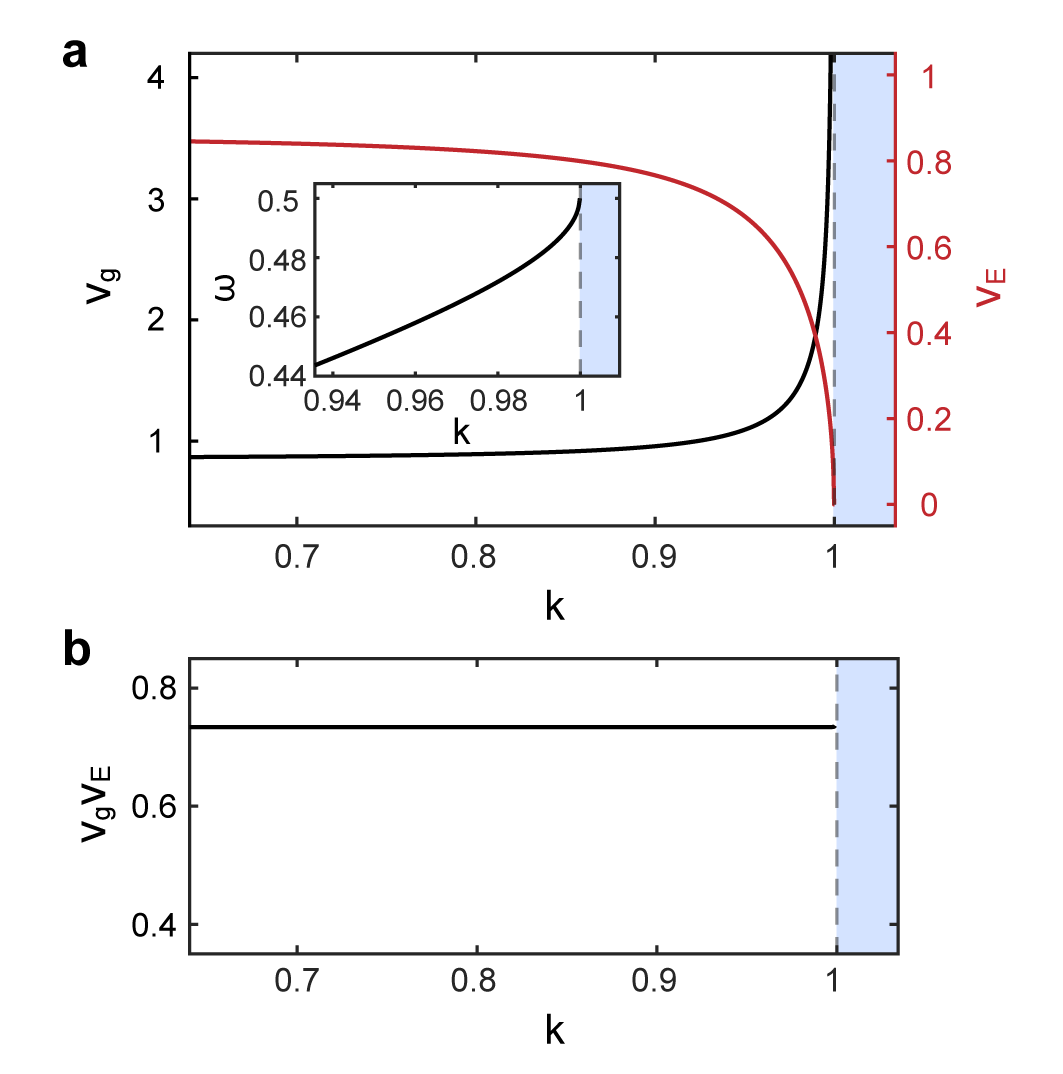}
  \caption{\label{fig:Fig1}Energy transport versus band-slope in a PTC. (a) Mode-resolved velocities on a passband: geometrical slope \(v_g=d\omega/dk\) (black) and energy-transport velocity \(v_E=\avgT{S_z}/\avgT{u}\) (red), both normalized by \(v_{\mathrm{ph,max}}\). The wavevector is shown in units of the momentum-gap edge \(k_c\) (vertical dashed line at \(k=1\)); the blue shading indicates the gap. Inset: Floquet band structure near the edge. 
  (b) Velocity-product law: $v_g v_E$ is constant throughout the passband; thus, as $k\to k_c$, the divergence of $v_g$ is exactly compensated by $v_E\to 0$.}
\end{figure}

\emph{From \(v_{\rm symp}\) to \(v_E\).—}
The ratio in Eq.~\eqref{eq:vsymp} is, in fact, the familiar energy-transport velocity.
For clarity we evaluate it in an \(n\)-harmonic truncation \(m\in\mathcal M\) (the identities hold in the full Floquet-harmonic representation).
In this basis, Eq.~\eqref{eq:harmonics} becomes the \(2n\times2n\) Maxwell--Floquet system
\begin{equation}
M(\omega,k)
=
\begin{pmatrix}
- K & \mu\Lambda\\[3pt]
- \Lambda\EpsMat & K
\end{pmatrix},
\label{eq:Mn}
\end{equation}
with \(K=k\,\mathbb I_n\), \(\Lambda=\mathrm{diag}(\{\omega_m\}_{m\in\mathcal M})\), and the Toeplitz matrix \((\EpsMat)_{mn}=\varepsilon_{m-n}\).
A direct calculation using Parseval's identity yields~\cite{SM}
\begin{equation}
\bra{R}\,\mathbb J\,\partial_k M\,\ket{R} = 4\,\avgT{S_z},
\quad
\bra{R}\,\mathbb J\,\partial_\omega M\,\ket{R} = -4\,\avgT{u},
\end{equation}
so that
\begin{equation}
v_{\rm symp}(k)
=
-\frac{\bra{R}\,\mathbb J\,\partial_k M\,\ket{R}}
{\bra{R}\,\mathbb J\,\partial_\omega M\,\ket{R}}
=
\frac{\avgT{S_z}}{\avgT{u}}
\equiv v_E(k).
\label{eq:vsymp=vE}
\end{equation}
 Thus the \(\mathbb J\)-weighted construction turns the Maxwell–Floquet eigenproblem into a transport diagnostic: on any simple real Floquet band, the physical velocity is \(v_E\), not the geometrical slope \(d\omega/dk\) (Fig.~\ref{fig:Fig1}). 
This distinction is sharp at momentum-gap edges. Near a branch-point edge \(k_c\), the two-harmonic model (and generically a simple Puiseux splitting~\cite{Kato1995Perturbation}) gives
\(\omega(k)\simeq \omega_0 \pm a\sqrt{\delta k}\) with \(\delta k=k-k_c\), hence \(d\omega/dk\propto |\delta k|^{-1/2}\).
By contrast, the flux-over-energy ratio is regular: typically \(\avgT{S_z}=O(\sqrt{|\delta k|})\) while \(\avgT{u}=O(1)\), so \(v_E\propto \sqrt{|\delta k|}\to 0\) (and \(\gamma\propto \sqrt{|\delta k|}\) inside the gap)~\cite{SM}. Equivalently, near a square-root edge the steepening of \(d\omega/dk\) is exactly paired with a collapse of the cycle-averaged flux, so \(v_g v_E\) stays \(O(1)\) even though \(v_g\) diverges (Fig.~\ref{fig:Fig1}). Furthermore, we prove analytically that the product of the group and energy
velocities is not only finite but exactly conserved across the passband~\cite{SM},
\begin{equation}
v_g\,v_E \;=\; \frac{1}{\mu}\left\langle \varepsilon^{-1}(t)\right\rangle_T \;=\;\langle v_{\rm ph}^2\rangle_T .
\label{eq:vgvE_conserved}
\end{equation}
Equation~\eqref{eq:vgvE_conserved} implies that the divergence of $v_g$ at the band edge is perfectly counterbalanced by the vanishing $v_E$, dictated solely by the temporal average of the inverse permittivity.

\emph{Momentum gaps and generalized velocity.—}
In a momentum gap the Floquet quasifrequency is complex, \(\omega=\omega_r+i\gamma\), so a single Floquet component \(e^{ikz-i\omega t}\) carries an overall envelope \(e^{\pm\gamma t}\).
To separate this trivial parametric amplification/attenuation from \emph{transport}, we write
\(u(z,t)=e^{2\gamma t}u_p(z,t)\) and \(S_z(z,t)=e^{2\gamma t}S_{z,p}(z,t)\),
where \(u_p\) and \(S_{z,p}\) are \(T\)-periodic for a single Floquet mode~\cite{SM}.
Equivalently, we introduce the weighted cycle average
\begin{equation}
\avgw{f}
\equiv \frac{1}{T}\int_0^T e^{-2\gamma t}\,f(t)\,dt,
\label{eq:weighted-average}
\end{equation}
which removes the envelope and is cycle-origin independent for a single mode.
We then define the generalized (gap) velocity
\begin{equation}
v_{\rm gen}(k)\equiv\frac{\avgw{S_z}}{\avgw{u}},
\label{eq:vgen-def}
\end{equation}
a flux-over-energy ratio that quantifies the net drift of energy in the presence of gain/loss. Using the same \(\mathbb J\)-weighted form as above, one finds the exact Maxwell–Floquet expression
\begin{equation}
v_{\rm gen}(k)
=
-\frac{\bra{R}\,\mathbb J\,\partial_k M\,\ket{R}}
      {\bra{R}\,\mathbb J\,\partial_\omega M\,\ket{R}},
\end{equation}
for the right Floquet vector \(\ket{R}\) satisfying \(M(\omega,k)\ket{R}=0\)~\cite{SM}.
By construction \(v_{\rm gen}\to v_E\) as \(\gamma\to0\), recovering the passband energy velocity. Furthermore, within the gap, the weighted cycle-averaged flux vanishes, $\avgw{S_z}=0$, while $\avgw{u}\neq0$, so Eq.~\eqref{eq:vgen-def} gives $v_{gen}=0$~\cite{SM}.

\emph{Energy centroid and wavepacket motion.—}
A wavepacket can exhibit rapid centroid motion in a PTC even when energy transport is slow, because temporal modulation continuously exchanges energy with the field.
To make this distinction explicit, we track the energy centroid.
For a localized, finite-energy wavepacket (so boundary terms vanish as \(|z|\to\infty\)), with energy density \(u(z,t)\) defined above, define
\(U(t)=\int u(z,t)\,dz\) and
\(Z(t)=\big[\int z\,u(z,t)\,dz\big]/U(t)\).
The quantity \(U(t)\) is the total electromagnetic energy of the wavepacket, while \(Z(t)\) is its center of energy.
Differentiating \(Z(t)\) and using the Poynting theorem yields the exact decomposition~\cite{SM}
\begin{equation}
\dot Z(t)=v_{\rm flux}(t)+v_{\rm mod}(t),
\label{eq:Zdot-general}
\end{equation}
where \(\dot Z(t)\) represents the velocity of this energy centroid.
The genuine transport contribution is
\begin{equation}
v_{\rm flux}(t) \equiv \frac{\int S_z(z,t)\,dz}{U(t)}.
\end{equation}
The second term, \(v_{\rm mod}(t)\), is \emph{not} a transport velocity; it is a modulation-driven centroid shift originating from the time-dependent weighting of the energy density.

To expose its structure, introduce the electric and magnetic energies and their fractions,
\begin{equation}
\begin{aligned}
U_E(t) &= \tfrac14\,\varepsilon(t)\!\int |E|^2\,dz, \qquad
U_H(t) = \tfrac14\,\mu\!\int |H|^2\,dz,\\
f_E(t) &= \frac{U_E(t)}{U(t)}, \qquad
f_H(t) = \frac{U_H(t)}{U(t)},
\end{aligned}
\end{equation}
as well as the corresponding intensity centroids
\begin{equation}
Z_E(t)=\frac{\int z|E|^2dz}{\int |E|^2dz},\qquad
Z_H(t)=\frac{\int z|H|^2dz}{\int |H|^2dz}.
\end{equation}
A short calculation gives~\cite{SM}
\begin{equation}
v_{\rm mod}(t)
=
-\frac{\dot\varepsilon(t)}{\varepsilon(t)}\,f_E(t)f_H(t)\,\bigl[Z_E(t)-Z_H(t)\bigr],
\label{eq:vmod-compact}
\end{equation}
with \(0\le f_Ef_H\le 1/4\).
Accordingly, a large \(v_{\rm mod}\) signals strong modulation-induced \emph{reweighting}
(i.e., energy exchange between the field and the drive, redistributing electric and magnetic
energy when \(Z_E\neq Z_H\)),
rather than rapid energy transport.

Viewed in \(k\)-space, where \(z\leftrightarrow i\partial_k\), the first spatial moment is generated by \(i\partial_k\).
Thus the centroids \(Z_E\) and \(Z_H\) contain, besides the band-slope term \(t\,\partial_k\omega\) (often denoted \(t\,v_g\)),
additional offsets controlled by the \(k\)-dependence of the periodic parts of the Floquet eigenmodes~\cite{SM}.
For a narrowband wavepacket centered at \(k_0\), the separation \(Z_E(t)-Z_H(t)\) reduces to a gauge-invariant mismatch
between the electric and magnetic \(k\)-space phase connections.
Equation~\eqref{eq:vmod-compact} then identifies \(v_{\rm mod}\) as a \emph{geometric drift}~\cite{SM}:
temporal modulation reweights the electric and magnetic energy fractions, and a non-zero electric--magnetic centroid separation
converts this reweighting into rapid centroid motion without implying rapid energy transport.

\begin{figure}[htb!]
  \centering
    \includegraphics[width=0.45\textwidth]{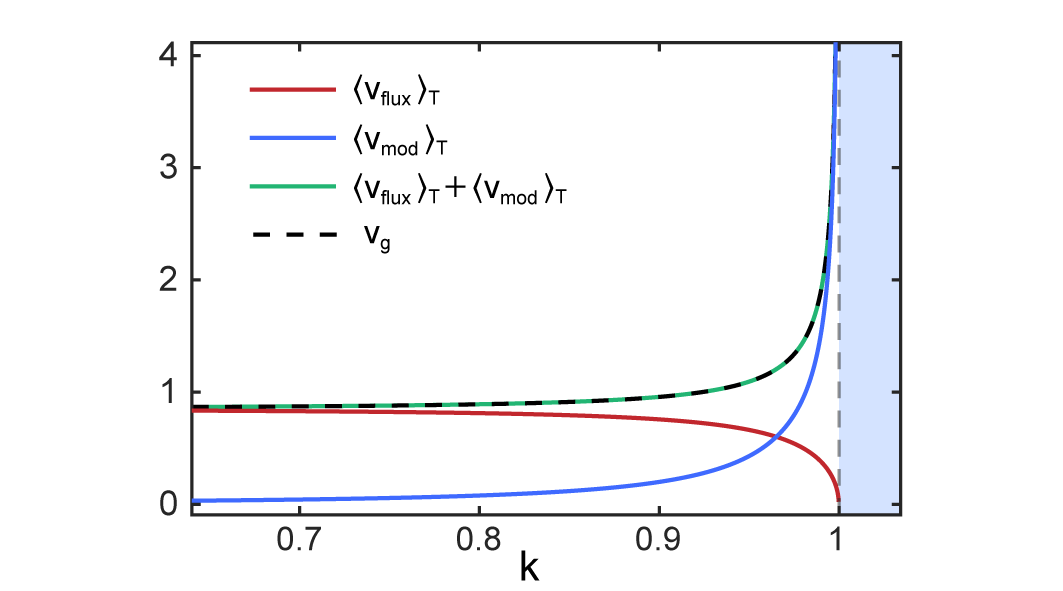}
  \caption{\label{fig:Fig2}Cycle-averaged centroid-velocity decomposition. In the narrowband limit \(\sigma_k\!\to\!0\), \(\avgT{\dot Z}=\avgT{v_{\rm flux}}+\avgT{v_{\rm mod}}\). Curves show \(\avgT{v_{\rm flux}}\) (red), \(\avgT{v_{\rm mod}}\) (blue), and their sum (green), compared with \(v_g=d\omega/dk\) (black dashed); all velocities are normalized by \(v_{\mathrm{ph,max}}\). The wavevector is normalized by the momentum-gap edge \(k_c\) (vertical dashed line at \(k=1\)); the blue shading indicates the gap.}
\end{figure}

To connect this instantaneous decomposition to the band picture, we now cycle-average over one modulation period. Cycle-averaging Eq.~\eqref{eq:Zdot-general} yields \(\avgT{\dot Z}=\avgT{v_{\rm flux}}+\avgT{v_{\rm mod}}\). For a spectrally narrow wavepacket on a passband, the cycle-averaged centroid drift is governed by the bounded transport part together with the modulation-driven drift~\cite{SM}. Figure~\ref{fig:Fig2} makes the key point transparent: as the momentum approaches a gap edge, the mode becomes increasingly standing-wave-like so that \(\avgT{v_{\rm flux}}\to0\), yet the geometrical group velocity \(v_g=d\omega/dk\) grows sharply. The resolution is that the steepening of \(v_g\) is carried by the modulation term: \(\avgT{v_{\rm flux}}+\avgT{v_{\rm mod}}\) tracks \(v_g\), whereas \(\avgT{v_{\rm flux}}\) alone remains bounded and vanishes at the edge. In other words, superluminal-looking group velocities or peak advances in a PTC are naturally interpreted as modulation-driven reweighting (distributed parametric pumping), not superluminal energy transport.

\emph{Discussion.—}
A PTC is not a passive medium: the modulation acts as a distributed pump that continually trades energy with the field.
This distinction is closely related to the standard lesson from fast-light phenomena in stationary dispersive media: an anomalous group velocity is not, by itself, a transport velocity. In both cases, superluminal-looking \(d\omega/dk\) must be separated from energy flow and does not imply a violation of causality. The origin of the anomaly, however, is different. Conventional fast-light effects arise from frequency-dependent dispersion in time-invariant media~\cite{garrett1970propagation, milonni2004fast,boyd2009controlling}, often accompanied by absorption or gain~\cite{wang2000gain,stenner2003speed}, whereas here the anomaly occurs in a time-periodically driven Floquet medium. Thus the PTC case should be viewed as a driven Floquet version of the group-velocity/transport- velocity distinction, rather than as ordinary fast light in a stationary dispersive medium.
In this setting, ``fast'' features in the Floquet band diagram often tell us less about how energy \emph{flows} than about how the drive \emph{reweights} the energy distribution between electric and magnetic sectors. Near a momentum-gap edge, the apparent divergence of the geometrical slope $d\omega/dk$ is carried by a modulation-induced \emph{geometric drift}: in a narrowband wavepacket the drive couples to a gauge-invariant mismatch between the electric and magnetic $k$-space phase connections (equivalently, an electric--magnetic centroid offset), shifting the center of energy even as the mode becomes standing-wave-like and the net flux collapses. Consistently, on any stable passband the exact velocity-product law
$ v_g v_E=\langle(\mu\varepsilon(t))^{-1}\rangle_T=\langle v_{\mathrm{ph}}^2\rangle_T $
pins the product of the group velocity and the cycle-averaged energy-transport velocity to a $k$-independent constant, so that the divergence of $v_g$ as $k\to k_c$ is necessarily compensated by $v_E\to 0$. 
The same transport collapse can be viewed in an equivalent Riemann-field representation~\cite{SM}. Defining the Riemann-field vector
\(\boldsymbol{\Phi}=\varepsilon^{1/4}
\big(\sqrt{\varepsilon}\,E+\sqrt{\mu}\,H,
-\sqrt{\varepsilon}\,E+\sqrt{\mu}\,H\big)^{T}\),
the Maxwell equations become
\begin{equation}
i\partial_t\boldsymbol{\Phi}
=
\left[
k v_{\rm ph}(t)\sigma_z
+
i m(t)\sigma_x
\right]\boldsymbol{\Phi},
\;\;
m(t)=\frac{\dot{\varepsilon}(t)}{4\varepsilon(t)} .
\end{equation}
This representation diagonalizes the Maxwell flux form: the cycle-averaged Poynting flux is controlled by
\(\boldsymbol{\Phi}_0^\dagger\sigma_z\boldsymbol{\Phi}_0\), while the energy density is controlled by a positive quadratic form. Hence the edge limit \(v_E\to0\) corresponds to a flux-form collapse,
\(\boldsymbol{\Phi}_0^\dagger\sigma_z\boldsymbol{\Phi}_0\to0\), not to a singular electromagnetic energy density.
This yields a practical reading rule for PTC band structures: steep dispersion primarily flags regimes dominated by pump-assisted reweighting, whereas transport is diagnosed by the energy-transport velocity. 
The same group-velocity/transport-velocity distinction should extend to other time-periodic wave systems, such as circuits or acoustic lattices, once the Poynting flux and electromagnetic energy density are replaced by the corresponding power flux and stored energy density.
In short, a PTC can move the \emph{center of energy} without transporting energy, and the energy-transport velocity---together with the conserved product law---tells the difference.

\section*{Methods}

\emph{Theoretical model and field convention.—}
We considered one-dimensional propagation in a spatially homogeneous, isotropic, source-free PTC with real, strictly positive periodic permittivity $\varepsilon(t+T)=\varepsilon(t)\ge \varepsilon_{\min}>0$ and constant permeability $\mu>0$. Unless stated otherwise, the medium was assumed to be nondispersive and lossless. With the complex-field convention used throughout, the electromagnetic energy density and Poynting flux were evaluated as
\begin{equation}
\begin{aligned}
&u(z,t)=\frac{1}{4}\left[\varepsilon(t)|E(z,t)|^2+\mu|H(z,t)|^2\right],
\\
&S_z(z,t)=\frac{1}{2}\mathrm{Re}\{E(z,t)H^*(z,t)\}.     
\end{aligned}   
\end{equation}

The period-averaged energy-transport velocity was defined as
\begin{equation}
v_E=\frac{\langle S_z\rangle_T}{\langle u\rangle_T},
\qquad
\langle f\rangle_T=\frac{1}{T}\int_0^T f(t)\,dt .
\end{equation}

\emph{Maxwell--Floquet eigenproblem.—}
The Floquet bands were calculated from the Maxwell--Floquet constraint
$M(\omega,k)|R\rangle=0$ derived in the main text. For each real wavevector
$k$, the fields were expanded in temporal Floquet harmonics,
\begin{equation}
\begin{aligned}
E(t)=\sum_{m\in\mathcal M} E_m e^{-i(\omega+m\Omega)t},\\
H(t)=\sum_{m\in\mathcal M} H_m e^{-i(\omega+m\Omega)t},    
\end{aligned}
\end{equation}
where $\Omega=2\pi/T$ and $\mathcal M$ is the retained set of harmonics.
The Floquet vector was written as $|R\rangle=(\Evec,\Hvec)^T$, with
$\Evec=\{E_m\}_{m\in\mathcal M}$ and
$\Hvec=\{H_m\}_{m\in\mathcal M}$. The dispersion relation was obtained from
the singularity condition of the Maxwell--Floquet matrix. Real quasifrequencies were identified as passbands, whereas complex
quasifrequencies were identified as momentum gaps.

\emph{Velocity calculation.—}
For each real Floquet branch, the group velocity was obtained from the slope of the quasifrequency band, $v_g=d\omega/dk$. The energy-transport velocity was evaluated from the period-averaged flux-over-energy ratio,
\begin{equation}
v_E=\frac{\langle S_z\rangle_T}{\langle u\rangle_T}.   
\end{equation}
On stable passbands, we further derived the analytical product law
\begin{equation}
v_g v_E=\left\langle\frac{1}{\mu\varepsilon(t)}\right\rangle_T ,    
\end{equation}
showing that the product of the group velocity and the energy-transport velocity is a $k$-independent constant fixed solely by the temporal
permittivity profile.

\emph{Wavepacket centroid analysis.—}
Localized finite-energy wavepackets were constructed by superposing modes from a single real Floquet branch with a narrow spectral envelope in $k$. Their energy-centroid dynamics were evaluated using the exact decomposition derived from the Poynting theorem,
\begin{equation}
\dot Z(t)=v_{\rm flux}(t)+v_{\rm mod}(t),    
\end{equation}
where $v_{\rm flux}$ is the total Poynting flux divided by the total energy, and $v_{\rm mod}$ is the modulation-induced contribution arising from the time-dependent reweighting of the electric and magnetic energy densities. For narrowband wavepackets, the electric and magnetic centroids were evaluated in $k$-space. In this limit, their separation reduces to the gauge-invariant mismatch between the corresponding phase connections.

\begin{acknowledgements}
\emph{Acknowledgements.—}
We thank J. B. Pendry for his critical reading of the manuscript and for insightful comments. This work is supported by the National Research Foundation of Korea (NRF) through the government of Korea (RS-2022-NR070636) and the Samsung Science and Technology Foundation (SSTF-BA240202). K.W.K. acknowledges financial support from the Basic Science Research Program through the NRF funded by the Ministry of Education (no. RS-2025-00521598) and the Korean Government (MSIT) (no. 2020R1A5A1016518). K.L. also acknowledges support from the KAIST Jang Young Sil Fellow Program.
\end{acknowledgements}

%

\clearpage
\onecolumngrid

{
\centering
\large \textbf{Supplemental Material: Energy Transport Velocity in Photonic Time Crystals} \\[2ex]

\normalsize Kyungmin Lee$^{1,*}$, Younsung Kim$^{1,*}$, Kun Woo Kim$^{2,\dagger}$, Bumki Min$^{1,\ddagger}$ \\[1ex]
\fontsize{9pt}{10pt}\selectfont
$^{1}$\textit{Department of Physics, Korea Advanced Institute of Science and Technology, Daejeon 34141, Republic of Korea} \\
$^{2}$\textit{Department of Physics, Chung-Ang University, 06974 Seoul, Republic of Korea} \\
\textit{$^{*}$These authors contributed equally to this work.} \\
\textit{$^{\dagger}$kunx@cau.ac.kr} \\
\textit{$^{\ddagger}$bmin@kaist.ac.kr}\\
}

\setcounter{secnumdepth}{3}
\setcounter{section}{0}
\renewcommand\thesection{Supplementary Note \arabic{section}}

\counterwithout{figure}{section}
\setcounter{figure}{0}
\renewcommand{\figurename}{Supplementary Figure}
\renewcommand\thefigure{\arabic{figure}}

\setcounter{table}{0}
\setcounter{equation}{0}
\counterwithin{table}{section}
\counterwithin{equation}{section}
\renewcommand\theequation{S\arabic{section}.\arabic{equation}}
\renewcommand\thetable{\thesection.\arabic{table}}

\section{\,\,\,\,\,\, Bound on the energy-transport velocity}
\label{sec:bound_vE}

\noindent In the main text we use the conventional complex-field representation, in which the quadratic quantities associated with the complex fields are understood as phasor-averaged real-field quantities. The rapidly oscillating phase factor is implicit in this notation. With this convention, the electromagnetic energy density and Poynting flux are
\begin{equation}
u(z,t) = \frac14\bigl(\varepsilon(t)|E(z,t)|^2 + \mu\,|H(z,t)|^2\bigr),
\qquad
S_z(z,t) = \frac12\Re\{E(z,t)\,H^*(z,t)\},
\label{eq:S1.uS_def}
\end{equation}
and they obey the Poynting theorem
\begin{equation}
\partial_t u + \partial_z S_z
= -\frac14\dot\varepsilon(t)|E(z,t)|^2.
\label{eq:S1.poynting}
\end{equation}
\subsection*{A. Derivation of the Poynting theorem}

For completeness, we derive Eqs.~\eqref{eq:S1.uS_def} and
\eqref{eq:S1.poynting}. We first consider the corresponding real
physical fields. In one dimension,
\begin{equation}
D_{\rm phys}(z,t)=\varepsilon(t)E_{\rm phys}(z,t),
\qquad
B_{\rm phys}(z,t)=\mu H_{\rm phys}(z,t),
\end{equation}
and Maxwell's equations read
\begin{equation}
\partial_z E_{\rm phys}
=
-\partial_t B_{\rm phys},
\qquad
\partial_z H_{\rm phys}
=
-\partial_t D_{\rm phys}.
\end{equation}
The instantaneous real-field Poynting flux is
\begin{equation}
S_{z,{\rm phys}}=E_{\rm phys}H_{\rm phys}.
\end{equation}
Taking its spatial derivative and using Maxwell's equations gives
\begin{align}
\partial_z S_{z,{\rm phys}}
&=
H_{\rm phys}\partial_zE_{\rm phys}
+
E_{\rm phys}\partial_zH_{\rm phys}
\nonumber\\
&=
-
\left[
E_{\rm phys}\partial_tD_{\rm phys}
+
H_{\rm phys}\partial_tB_{\rm phys}
\right].
\end{align}
We now rewrite the term in brackets using
\(D_{\rm phys}=\varepsilon(t)E_{\rm phys}\) and
\(B_{\rm phys}=\mu H_{\rm phys}\):
\begin{align}
E_{\rm phys}\partial_tD_{\rm phys}
+
H_{\rm phys}\partial_tB_{\rm phys}
&=
E_{\rm phys}\partial_t[\varepsilon(t)E_{\rm phys}]
+
\mu H_{\rm phys}\partial_tH_{\rm phys}
\nonumber\\
&=
\partial_t\left[
\frac12\varepsilon(t)E_{\rm phys}^2
+
\frac12\mu H_{\rm phys}^2
\right]
+
\frac12\dot{\varepsilon}(t)E_{\rm phys}^2 .
\end{align}
Thus, with the real-field energy density
\begin{equation}
u_{\rm phys}
=
\frac12\left[
\varepsilon(t)E_{\rm phys}^2+\mu H_{\rm phys}^2
\right],
\end{equation}
we obtain
\begin{equation}
E_{\rm phys}\partial_tD_{\rm phys}
+
H_{\rm phys}\partial_tB_{\rm phys}
=
\partial_tu_{\rm phys}
+
\frac12\dot{\varepsilon}(t)E_{\rm phys}^2 .
\end{equation}
Combining this result with the expression for
\(\partial_z S_{z,{\rm phys}}\) yields the real-field energy balance
\begin{equation}
\partial_tu_{\rm phys}
+
\partial_zS_{z,{\rm phys}}
=
-\frac12\dot{\varepsilon}(t)E_{\rm phys}^2 .
\end{equation}

For the complex fields used in Eq.~\eqref{eq:S1.uS_def}, the quadratic quantities are the corresponding phasor-averaged real-field quantities. The overbar denotes the time average over the fast oscillation implicit in the complex-field representation, and should be distinguished from the modulation-period average \(\langle\cdot\rangle_T\) introduced below. Explicitly,
\begin{equation}
\overline{E_{\rm phys}^2}
=
\frac12|E|^2,
\qquad
\overline{H_{\rm phys}^2}
=
\frac12|H|^2,
\end{equation}
and
\begin{equation}
\overline{E_{\rm phys}H_{\rm phys}}
=
\frac12\Re\{EH^*\}.
\end{equation}
Applying this phasor average to the real-field energy density and flux
gives
\begin{equation}
u(z,t)
=
\frac14\left[
\varepsilon(t)|E(z,t)|^2+\mu |H(z,t)|^2
\right],
\qquad
S_z(z,t)
=
\frac12\Re\{E(z,t)H^*(z,t)\},
\end{equation}
which is Eq.~\eqref{eq:S1.uS_def}. Applying the same phasor average to the real-field energy balance gives
\begin{equation}
\partial_tu+\partial_zS_z
=
-\frac14\dot{\varepsilon}(t)|E(z,t)|^2,
\end{equation}
which is Eq.~\eqref{eq:S1.poynting}.

The cycle-averaged energy density and flux over one modulation period \(T\) are
\begin{equation}
\langle u\rangle_T
= \frac{1}{T}\int_0^T u(z,t)\,dt,
\qquad
\langle S_z\rangle_T
= \frac{1}{T}\int_0^T S_z(z,t)\,dt,
\label{eq:S1.uS_avg}
\end{equation}
and we define
\begin{equation}
v_E \equiv \frac{\langle S_z\rangle_T}{\langle u\rangle_T}.
\label{eq:S1.vE_def}
\end{equation}
Here \(\varepsilon(t+T)=\varepsilon(t)\ge\varepsilon_{\min}>0\) is real and strictly positive, and \(\mu>0\) is constant.
In this Note we justify that the period averages are finite on real Floquet bands and derive the bound
\(|v_E|\le v_{\rm ph,max}\equiv 1/\sqrt{\mu\,\varepsilon_{\min}}\).
In SI units, with \(\varepsilon(t)=\varepsilon_v\varepsilon_r(t)\) and \(\mu=\mu_v\mu_r\), this can be written as
\(v_{\rm ph,max}=c/\sqrt{\mu_r\,\varepsilon_{r,\min}}\le c\) provided \(\mu\,\varepsilon_{\min}\ge \mu_v\varepsilon_v\) (equivalently \(\mu_r\,\varepsilon_{r,\min}\ge 1\)).

\subsection{Floquet modes and finiteness of the cycle averages}

\noindent On a simple real Floquet band, it is convenient to write the fields as
\begin{equation}
E(z,t)=e^{ikz-i\omega t}\,\mathcal E(t),\qquad
H(z,t)=e^{ikz-i\omega t}\,\mathcal H(t),
\end{equation}
where \(\mathcal E(t)\) and \(\mathcal H(t)\) are \(T\)-periodic and \(\omega\in\mathbb{R}\).
Because \(\mathcal E(t)\), \(\mathcal H(t)\), and \(\varepsilon(t)\) are bounded and \(T\)-periodic with \(\varepsilon(t)\ge\varepsilon_{\min}>0\), the quantities \(u(z,t)\) and \(S_z(z,t)\) are bounded (hence integrable) on $[0,T]$. Thus the period averages exist and are finite. Moreover, for any nontrivial mode $u\ge 0$ and $u\not\equiv 0$, implying $\langle u\rangle_T>0$ and thus $v_E$ is well defined.

\subsection{Pointwise inequality and the phase-velocity bound}

\noindent From Eq.~\eqref{eq:S1.uS_def} we have
\begin{equation}
|S_z| = \frac12|\Re(EH^*)|
\le \frac12|E|\,|H|.
\label{eq:S1.Sz_bound1}
\end{equation}
Using the arithmetic--geometric mean inequality for the nonnegative numbers
\(\varepsilon\,|E|^2\) and \(\mu\,|H|^2\),
\begin{equation}
2\sqrt{\varepsilon\,\mu}\,|E|\,|H|
\le \varepsilon\,|E|^2 + \mu\,|H|^2,
\label{eq:S1.AGM}
\end{equation}
we obtain
\begin{equation}
\frac12|E|\,|H|
\le \frac{\varepsilon\,|E|^2 + \mu\,|H|^2}{4\sqrt{\varepsilon\,\mu}}
= \frac{u}{\sqrt{\varepsilon\,\mu}},
\end{equation}
where we used the definition of \(u\) in Eq.~\eqref{eq:S1.uS_def}.
Combining with Eq.~\eqref{eq:S1.Sz_bound1} yields the pointwise inequality
\begin{equation}
|S_z(z,t)|
\le \frac{u(z,t)}{\sqrt{\varepsilon(t)\,\mu}}.
\label{eq:S1.Sz_pointwise}
\end{equation}
Since \(\varepsilon(t)\ge\varepsilon_{\min}>0\) by assumption, we further have
\begin{equation}
\frac{1}{\sqrt{\varepsilon(t)\,\mu}} \le \frac{1}{\sqrt{\varepsilon_{\min}\,\mu}}
\equiv v_{\rm ph,max},
\label{eq:S1.vph_max}
\end{equation}
where \(v_{\rm ph,max}\) is the largest instantaneous phase velocity of the medium.

\subsection{Bound on the cycle-averaged energy velocity}

\noindent From Eq.~\eqref{eq:S1.vE_def} and using $\langle u\rangle_T>0$ for any nontrivial mode, we have
\begin{equation}
|v_E|
= \left|\frac{\langle S_z\rangle_T}{\langle u\rangle_T}\right|
\le \frac{\langle |S_z|\rangle_T}{\langle u\rangle_T},
\end{equation}
where we used \(|\langle S_z\rangle_T|\le \langle|S_z|\rangle_T\).
Averaging the pointwise inequality~\eqref{eq:S1.Sz_pointwise} over one period gives
\begin{equation}
\langle |S_z|\rangle_T
\le \left\langle \frac{u}{\sqrt{\varepsilon(t)\,\mu}}\right\rangle_T.
\end{equation}
Using \(\varepsilon(t)\ge\varepsilon_{\min}\), we obtain
\begin{equation}
\frac{u(z,t)}{\sqrt{\varepsilon(t)\,\mu}}
\le \frac{u(z,t)}{\sqrt{\varepsilon_{\min}\,\mu}}
= v_{\rm ph,max}\,u(z,t),
\end{equation}
so that
\begin{equation}
\left\langle \frac{u}{\sqrt{\varepsilon(t)\,\mu}}\right\rangle_T
\le v_{\rm ph,max}\,\langle u\rangle_T.
\end{equation}
Combining these yields
\begin{equation}
|v_E|\le v_{\rm ph,max}=\frac{1}{\sqrt{\mu\,\varepsilon_{\min}}}.
\end{equation}

\section{\,\,\,\,\,\, FLUX AND ENERGY RELATIONS IN THE FLOQUET-HARMONIC REPRESENTATION}
\label{sec:bilinears}

\noindent In the main text we state that
\begin{equation}
\langle R|\mathbb J\,\partial_k M|R\rangle = 4\,\langle S_z\rangle_T,
\qquad
\langle R|\mathbb J\,\partial_\omega M|R\rangle = -4\,\langle u\rangle_T,
\label{eq:S2.claim}
\end{equation}
where \(\langle S_z\rangle_T\) and \(\langle u\rangle_T\) are the cycle-averaged flux and energy density defined in~\ref{sec:bound_vE}.
Here we provide a detailed derivation.

\subsection{Derivation of the band-slope formula}

For completeness, we derive the band-slope formula used in the main
text. The Maxwell--Floquet matrix equation is
\begin{equation}
M(\omega,k)|R\rangle=0 .
\end{equation}
On a simple real band, the quasifrequency can be regarded as a function of real wavevector, $\omega=\omega(k)$, and the right null vector as
$|R\rangle=|R(k)\rangle$. Differentiating with respect to $k$ gives
\begin{equation}
\left(\partial_k M + \frac{d\omega}{dk}\partial_\omega M\right)|R\rangle
+ M\frac{d|R\rangle}{dk}=0 .
\end{equation}
Introducing the left null vector $\langle L|$, defined by
\begin{equation}
\langle L|M(\omega,k)=0 ,
\end{equation}
and multiplying the differentiated equation from the left by
$\langle L|$, the last term vanishes. We therefore obtain
\begin{equation}
\langle L|\partial_k M|R\rangle
+
\frac{d\omega}{dk}
\langle L|\partial_\omega M|R\rangle
=0 .
\end{equation}
Provided that
$\langle L|\partial_\omega M|R\rangle\neq 0$, which holds away from
the momentum-gap edges, the band slope is
\begin{equation}
v_g(k)\equiv\frac{d\omega}{dk}
=
-\frac{\langle L|\partial_k M|R\rangle}
{\langle L|\partial_\omega M|R\rangle}.
\end{equation}
This is Eq.~(4) of the main text. It gives the geometrical slope of the Floquet dispersion and should be distinguished from the ratio introduced below, which gives the energy-transport velocity.

\subsection{Matrix structure and symplectic pairing}

\noindent We work in the Floquet-harmonic representation spanned by a retained set of harmonics
$m\in\mathcal M\subset\mathbb Z$ with $|\mathcal M|=n$ (in the main text we use the two-harmonic choice $\mathcal M=\{0,-1\}$, hence $n=2$).
In this $n$-harmonic basis the Maxwell--Floquet matrix takes the block form
\begin{equation}
M(\omega,k)
=
\begin{pmatrix}
- K & \mu\Lambda\\[3pt]
- \Lambda\EpsMat & K
\end{pmatrix},
\label{eq:S2.M2}
\end{equation}
with $K\equiv k\,\mathbb I_n$,
$\Lambda=\mathrm{diag}(\{\omega_m\}_{m\in\mathcal M})$ where $\omega_m=\omega+m\Omega$,
and $(\EpsMat)_{mn}\equiv \varepsilon_{m-n}$ for $m,n\in\mathcal M$; i.e., $\EpsMat$ is the Toeplitz matrix built from the Fourier coefficients $\varepsilon_\ell$.
The right null vector is
\begin{equation}
|R\rangle =
\begin{pmatrix}
\bm{E}\\ \bm{H}
\end{pmatrix},
\qquad
\bm{E}=\begin{pmatrix} \{E_m\}_{m\in\mathcal M}\end{pmatrix}^{\mathsf T},
\quad
\bm{H}=\begin{pmatrix} \{H_m\}_{m\in\mathcal M}\end{pmatrix}^{\mathsf T},
\label{eq:S2.Rvec}
\end{equation}
and the symplectic matrix is
\begin{equation}
\mathbb J =
\begin{pmatrix}
0 & \mathbb I_n\\[2pt]
-\mathbb I_n & 0
\end{pmatrix}.
\label{eq:S2.Jdef}
\end{equation}

\noindent
Throughout this Note we use the standard Hermitian pairing in Sambe space, so bras denote Hermitian adjoints, $\langle A|\equiv(|A\rangle)^\dagger$.
For a lossless medium with real $\varepsilon(t)$, we evaluate the $\mathbb J$-weighted forms in Eq.~\eqref{eq:S2.claim} under this pairing.
In particular,
\begin{equation}
\langle R|\mathbb J
= \bigl(\bm{E}^\dagger, \bm{H}^\dagger\bigr)
\begin{pmatrix}
0 & \mathbb I_n\\[2pt]
-\mathbb I_n & 0
\end{pmatrix}
= \bigl(-\bm{H}^\dagger,\,\bm{E}^\dagger\bigr).
\label{eq:S2.Ldef}
\end{equation}
In this basis the derivatives of $M$ read
\begin{equation}
\partial_k M =
\begin{pmatrix}
-\mathbb I_n & 0\\
0 & \mathbb I_n
\end{pmatrix},
\qquad
\partial_\omega M =
\begin{pmatrix}
0 & \mu\mathbb I_n\\[2pt]
-\EpsMat & 0
\end{pmatrix}.
\label{eq:S2.dM}
\end{equation}

\subsection{Flux term and cycle-averaged Poynting flux}

\noindent We first evaluate the numerator \(\langle R|\mathbb J\partial_k M|R\rangle\).
Using Eqs.~\eqref{eq:S2.Jdef} and \eqref{eq:S2.dM} we compute
\begin{equation}
\mathbb J\,\partial_k M
=
\begin{pmatrix}
0 & \mathbb I_n\\[2pt]
-\mathbb I_n & 0
\end{pmatrix}
\begin{pmatrix}
-\mathbb I_n & 0\\
0 & \mathbb I_n
\end{pmatrix}
=
\begin{pmatrix}
0 & \mathbb I_n\\[2pt]
\mathbb I_n & 0
\end{pmatrix}.
\label{eq:S2.JdMk}
\end{equation}
Therefore
\begin{equation}
\langle R|\mathbb J\partial_k M|R\rangle
= \bigl(\bm{E}^\dagger,\bm{H}^\dagger\bigr)
\begin{pmatrix}
0 & \mathbb I_n\\[2pt]
\mathbb I_n & 0
\end{pmatrix}
\begin{pmatrix}
\bm{E}\\ \bm{H}
\end{pmatrix}
= \bm{E}^\dagger \bm{H} + \bm{H}^\dagger \bm{E}.
\label{eq:S2.num_EH}
\end{equation}
On the other hand, the Floquet-periodic parts admit the harmonic expansions
\begin{equation}
\mathcal E(t) = \sum_{m\in\mathcal M} E_m e^{-im\Omega t},\qquad
\mathcal H(t) = \sum_{m\in\mathcal M} H_m e^{-im\Omega t},
\label{eq:S2.Floquet_EH}
\end{equation}
so that the period average of the mixed term is
\begin{equation}
\frac{1}{T}\int_0^T \mathcal E(t)\,\mathcal H^*(t)\,dt
= \sum_{m\in\mathcal M} E_m H_m^*.
\label{eq:S2.Parseval_EH}
\end{equation}
Using the definition of the cycle-averaged flux,
\begin{equation}
\langle S_z\rangle_T
= \frac{1}{T}\int_0^T \frac12\Re\{\mathcal E(t)\,\mathcal H^*(t)\}\,dt
= \frac12\Re\left[\frac{1}{T}\int_0^T \mathcal E(t)\,\mathcal H^*(t)\,dt\right]
= \frac12\Re\Bigl(\sum_{m\in\mathcal M} E_m H_m^*\Bigr),
\label{eq:S2.Savg_def}
\end{equation}
we can rewrite Eq.~\eqref{eq:S2.Savg_def} in terms of the vectors \(\bm{E}=(\{E_m\})^{\mathsf T}\), \(\bm{H}=(\{H_m\})^{\mathsf T}\):
\begin{equation}
\langle S_z\rangle_T
= \frac12\Re\bigl(\bm{E}^\dagger \bm{H}\bigr)
= \frac{\bm{E}^\dagger \bm{H} + \bm{H}^\dagger \bm{E}}{4}.
\label{eq:S2.Savg_EH}
\end{equation}
Combining Eq.~\eqref{eq:S2.num_EH} with Eq.~\eqref{eq:S2.Savg_EH}, we obtain
\begin{equation}
\langle R| \mathbb J\partial_k M|R\rangle
= \bm{E}^\dagger \bm{H} + \bm{H}^\dagger \bm{E}
= 4\,\langle S_z\rangle_T,
\label{eq:S2.num_final}
\end{equation}
which is the first relation in Eq.~\eqref{eq:S2.claim}.

\subsection{Energy term and cycle-averaged energy density}

\noindent We now turn to the denominator \(\langle R| \mathbb J \partial_\omega M|R\rangle\).
From Eqs.~\eqref{eq:S2.Jdef} and \eqref{eq:S2.dM},
\begin{equation}
\mathbb J\,\partial_\omega M
=
\begin{pmatrix}
0 & \mathbb I_n\\[2pt]
-\mathbb I_n & 0
\end{pmatrix}
\begin{pmatrix}
0 & \mu\mathbb I_n\\[2pt]
-\EpsMat & 0
\end{pmatrix}
=
\begin{pmatrix}
-\EpsMat & 0\\[2pt]
0 & -\mu\mathbb I_n
\end{pmatrix}.
\label{eq:S2.JdMomega}
\end{equation}
Therefore
\begin{equation}
\langle R|\mathbb J\partial_\omega M|R\rangle
= \bigl(\bm{E}^\dagger,\bm{H}^\dagger\bigr)
\begin{pmatrix}
-\EpsMat & 0\\[2pt]
0 & -\mu\mathbb I_n
\end{pmatrix}
\begin{pmatrix}
\bm{E}\\ \bm{H}
\end{pmatrix}
= -\bigl(\bm{E}^\dagger\EpsMat \bm{E} + \mu\, \bm{H}^\dagger \bm{H}\bigr).
\label{eq:S2.den_EH}
\end{equation}
Using the harmonic expansions and Parseval’s identity, the period averages are
\begin{equation}
\frac{1}{T}\int_0^T \varepsilon(t)\,|\mathcal E(t)|^2\,dt
= \bm{E}^\dagger \EpsMat \bm{E},
\qquad
\frac{1}{T}\int_0^T \mu\,|\mathcal H(t)|^2\,dt
= \mu\,\bm{H}^\dagger \bm{H},
\label{eq:S2.Parseval_u}
\end{equation}
where \(\EpsMat\) is the Toeplitz matrix \([\varepsilon_{m-n}]\) restricted to the chosen Floquet-harmonic basis. With the energy-density definition~\eqref{eq:S1.uS_def},
\begin{equation}
\langle u\rangle_T
= \frac{1}{T}\int_0^T \frac14\bigl(\varepsilon|\mathcal E|^2+\mu|\mathcal H|^2\bigr)\,dt
= \frac14\bigl(\bm{E}^\dagger\EpsMat \bm{E} + \mu\, \bm{H}^\dagger \bm{H}\bigr),
\label{eq:S2.uavg_EH}
\end{equation}
we immediately obtain
\begin{equation}
\bm{E}^\dagger\EpsMat \bm{E} + \mu\, \bm{H}^\dagger \bm{H}
= 4\,\langle u\rangle_T.
\label{eq:S2.uavg_rel}
\end{equation}
Combining Eq.~\eqref{eq:S2.den_EH} with Eq.~\eqref{eq:S2.uavg_rel}, we arrive at
\begin{equation}
\langle R|\mathbb J\partial_\omega M|R\rangle
= -\bigl(\bm{E}^\dagger\EpsMat \bm{E} + \mu\, \bm{H}^\dagger \bm{H}\bigr)
= -4\,\langle u\rangle_T,
\label{eq:S2.den_final}
\end{equation}
which is the second relation in Eq.~\eqref{eq:S2.claim}.

\section{\,\,\,\,\,\, Gap-edge scaling and a two-harmonic benchmark for $v_g$ and $v_E$}
\label{sec:edge_scaling_twoharm_benchmark}

\noindent
\ref{sec:bilinears} establishes the exact $\mathbb J$-weighted identities
\begin{equation}
\langle R|\mathbb J\,\partial_k M|R\rangle = 4\,\langle S_z\rangle_T,
\qquad
\langle R|\mathbb J\,\partial_\omega M|R\rangle = -4\,\langle u\rangle_T,
\label{eq:Sx.claim}
\end{equation}
which imply
\begin{equation}
v_E(k)\equiv \frac{\langle S_z\rangle_T}{\langle u\rangle_T}
=
-\frac{\langle R|\mathbb J\partial_kM|R\rangle}{\langle R|\mathbb J\partial_\omega M|R\rangle}.
\label{eq:Sx.vE_bilinear}
\end{equation}
In this Note we (i) collect the standard Puiseux (square-root) scaling near a momentum-gap edge and the resulting finiteness of $v_g v_E$, and (ii) provide an explicit two-harmonic benchmark: a closed-form dispersion equation and compact implicit expressions for $v_g=d\omega/dk$ and $v_E$, useful for analytic checks and for clarifying how the edge scaling emerges in the minimal model.

\subsection{Scaling near a momentum-gap edge}
\label{subsec:Sx_edge_scaling}

\noindent
A Floquet band is defined by the singularity condition
\begin{equation}
M(\omega,k)\,|R\rangle = 0,
\label{eq:Sx_MR0}
\end{equation}
or equivalently, in an $n$-harmonic truncation of the Floquet-harmonic representation, by the scalar dispersion equation
\begin{equation}
F(\omega,k) \equiv \det M(\omega,k)=0.
\label{eq:Sx_Fdef}
\end{equation}
We consider a momentum-gap edge $(\omega_0,k_c)$ that is a \emph{simple (square-root) branch point}, i.e., a point where $F$ has a double root in $\omega$,
\begin{equation}
F(\omega_0,k_c)=0,\qquad \partial_\omega F(\omega_0,k_c)=0,
\label{eq:Sx_double_root}
\end{equation}
and we assume the nondegeneracy conditions
\begin{equation}
\partial_{\omega\omega}F(\omega_0,k_c)\neq 0,\qquad \partial_kF(\omega_0,k_c)\neq 0,
\label{eq:Sx_nondeg}
\end{equation}
which ensure that the local splitting is of Puiseux (square-root) form~\cite{SM-Kato1995Perturbation}. Writing $\delta\omega\equiv \omega-\omega_0$ and $\delta k\equiv k-k_c$, the lowest-order Taylor expansion gives
\begin{equation}
0=F(\omega,k)\simeq \frac{1}{2}\big(\partial_{\omega\omega}F\big)_0\,(\delta\omega)^2
+\big(\partial_kF\big)_0\,\delta k,
\label{eq:Sx_Taylor}
\end{equation}
hence the Puiseux splitting
\begin{equation}
\delta\omega \simeq \pm a\sqrt{\delta k},
\qquad
a\equiv \sqrt{-\frac{2(\partial_kF)_0}{(\partial_{\omega\omega}F)_0}},
\label{eq:Sx_puiseux}
\end{equation}
and the divergent band slope $d\omega/dk\propto |\delta k|^{-1/2}$ on the passband side.
For a lossless medium with real $\varepsilon(t)$, $F(\omega,k)$ has real coefficients for real $(\omega,k)$, so inside the gap the two local branches form a complex-conjugate pair, $\omega=\omega_r\pm i\gamma$, giving
\begin{equation}
\gamma \equiv |\Im\,\omega| \propto \sqrt{|\delta k|}.
\label{eq:Sx_gamma_scaling}
\end{equation}
\noindent
In contrast, the flux-over-energy diagnostic remains regular because it is governed by the \(\mathbb J\)-weighted identities:
\begin{equation}
\langle R|\,\mathbb J\,\partial_k M\,|R\rangle = 4\langle S_z\rangle_T,
\qquad
\langle R|\,\mathbb J\,\partial_\omega M\,|R\rangle = -4\langle u\rangle_T,
\label{eq:Sx_flux_energy_bilinears}
\end{equation}
so that
\begin{equation}
v_E \equiv \frac{\langle S_z\rangle_T}{\langle u\rangle_T}
= -\frac{\langle R|\mathbb J\partial_kM|R\rangle}{\langle R|\mathbb J\partial_\omega M|R\rangle}.
\label{eq:Sx_vE_ratio}
\end{equation}
At a \emph{symmetry-selected} gap edge in a lossless PTC the coalescing mode is standing-wave-like, hence the net cycle-averaged flux vanishes:
\begin{equation}
\langle S_z\rangle_T(k_c)=0
\quad\Leftrightarrow\quad
\langle R_0|\mathbb J\partial_kM|R_0\rangle=0,
\qquad |R_0\rangle\equiv |R(k_c)\rangle.
\label{eq:Sx_flux_zero_edge}
\end{equation}
To control the eigenvector expansion, we fix a normalization of $|R(k)\rangle$ such that the cycle-averaged energy remains finite and nonzero at the edge, $\langle u\rangle_T(k_c)>0$ (equivalently, $\langle R_0|\mathbb J\partial_\omega M|R_0\rangle=-4\langle u\rangle_T(k_c)\neq 0$).
With this convention, near a simple branch point the right null vector admits a Puiseux expansion
\begin{equation}
|R(k)\rangle = |R_0\rangle + \sqrt{\delta k}\,|R_1\rangle + O(\delta k),
\label{eq:Sx_puiseux_vec}
\end{equation}
where $\sqrt{\delta k}$ is taken on the passband side ($\delta k\to 0^+$). Substituting Eq.~\eqref{eq:Sx_puiseux_vec} into the first identity in Eq.~\eqref{eq:Sx_flux_energy_bilinears} yields
\begin{equation}
\langle R|\mathbb J\partial_kM|R\rangle
=
\underbrace{\langle R_0|\mathbb J\partial_kM|R_0\rangle}_{=\,0}
+2\,\Re\!\left[\langle R_0|\mathbb J\partial_kM|R_1\rangle\right]\sqrt{\delta k}
+O(\delta k),
\label{eq:Sx_flux_expand}
\end{equation}
so that
\begin{equation}
\langle S_z\rangle_T = O(\sqrt{|\delta k|}).
\label{eq:Sx_Sz_scaling}
\end{equation}
Meanwhile, for any nontrivial lossless mode the cycle-averaged energy density remains finite at the edge,
\begin{equation}
\langle u\rangle_T = O(1),
\label{eq:Sx_u_scaling}
\end{equation}
and therefore, at a symmetry-selected edge where Eq.~\eqref{eq:Sx_flux_zero_edge} holds,
\begin{equation}
v_E=\frac{\langle S_z\rangle_T}{\langle u\rangle_T}
=O(\sqrt{|\delta k|})\xrightarrow[\delta k\to 0]{}0,
\label{eq:Sx_vE_scaling}
\end{equation}
even though the band slope diverges as $d\omega/dk\propto |\delta k|^{-1/2}$.

\subsection{Finite product of the band slope and the transport velocity}
\label{subsec:Sx_edge_product}

\noindent
The complementary edge scalings obtained above imply that the product of the geometrical band slope and the energy-transport velocity remains finite at a simple (square-root) gap edge. From the Puiseux splitting in Eq.~\eqref{eq:Sx_puiseux}, for $\delta k\to 0^+$ on the passband side we have
\begin{equation}
v_g(k)\equiv \frac{d\omega}{dk}
=\pm \frac{a}{2\sqrt{\delta k}} + O(1).
\label{eq:Sx_vg_edge}
\end{equation}
Meanwhile, using Eq.~\eqref{eq:Sx_puiseux_vec} together with the flux bilinear in Eq.~\eqref{eq:Sx_flux_energy_bilinears} and the standing-wave condition~\eqref{eq:Sx_flux_zero_edge}, the cycle-averaged flux admits the coefficient-level expansion
\begin{equation}
\langle S_z\rangle_T
=\frac12\Re\!\left[\langle R_0|\mathbb J\partial_kM|R_1\rangle\right]\sqrt{\delta k}
+O(\delta k).
\label{eq:Sx_Sz_edge_coeff}
\end{equation}
For the denominator, the energy bilinear is continuous at the edge (under the normalization fixed above):
\begin{equation}
\langle R|\mathbb J\partial_\omega M|R\rangle
=
\langle R_0|\mathbb J\partial_\omega M|R_0\rangle + O(\sqrt{\delta k})
= -4\langle u\rangle_T(k_c) + O(\sqrt{\delta k}),
\label{eq:Sx_energy_cont}
\end{equation}
so that $\langle u\rangle_T=\langle u\rangle_T(k_c)+O(\sqrt{\delta k})$ with $\langle u\rangle_T(k_c)>0$.
Defining the (edge) coefficient
\begin{equation}
C_E \equiv \frac{\frac12\Re\!\left[\langle R_0|\mathbb J\partial_kM|R_1\rangle\right]}{\langle u\rangle_T(k_c)},
\label{eq:Sx_CE_def}
\end{equation}
we obtain
\begin{equation}
v_E(k)=\frac{\langle S_z\rangle_T}{\langle u\rangle_T}
= C_E\,\sqrt{\delta k}+O(\delta k).
\label{eq:Sx_vE_edge_coeff}
\end{equation}
Combining Eqs.~\eqref{eq:Sx_vg_edge} and \eqref{eq:Sx_vE_edge_coeff} yields
\begin{equation}
v_g(k)\,v_E(k)
=
\pm \frac{a\,C_E}{2}+O(\sqrt{\delta k}),
\label{eq:Sx_vgvE_finite}
\end{equation}
so the product remains finite as $k\to k_c$ for a generic square-root edge satisfying Eq.~\eqref{eq:Sx_flux_zero_edge}.
The limiting value in Eq.~\eqref{eq:Sx_vgvE_finite} is not universal: while $a$ is fixed by derivatives of $F(\omega,k)=\det M(\omega,k)$ at the branch point [Eq.~\eqref{eq:Sx_puiseux}], the coefficient $C_E$ depends on the
mode-mixing data through $|R_1\rangle$.
We note that $C_E$ is nevertheless well defined: $|R_1\rangle$ is defined only up to $|R_1\rangle\to |R_1\rangle+\alpha|R_0\rangle$, but Eq.~\eqref{eq:Sx_flux_zero_edge} ensures that $\Re[\langle R_0|\mathbb J\partial_kM|R_0\rangle]=0$, so $C_E$ is invariant under this freedom.

\subsection{Two-harmonic truncation example: implicit closed forms for $F_2$, $v_g$, and $v_E$}
\label{subsec:Sx_twoharm_benchmark}

\noindent
We now specialize to the two-harmonic truncation $m\in\{0,-1\}$ (so $\omega_{m=0}=\omega$ and $\omega_{m=-1}=\omega-\Omega$).
From the harmonic relations $kE_m=\mu\omega_mH_m$ and $kH_m=\omega_m\sum_n\varepsilon_{m-n}E_n$, eliminating $H_m$ gives
\begin{equation}
\frac{k^2}{\mu}E_m=\omega_m^2\sum_{n\in\{0,-1\}}\varepsilon_{m-n}E_n.
\end{equation}
Defining $A\equiv k^2/\mu$, this yields the $2\times2$ system
\begin{equation}
\begin{pmatrix}
A-\varepsilon_0\omega^2 & -\varepsilon_{+1}\omega^2\\[2pt]
-\varepsilon_{-1}(\omega-\Omega)^2 & A-\varepsilon_0(\omega-\Omega)^2
\end{pmatrix}
\begin{pmatrix}
E_0\\ E_{-1}
\end{pmatrix}=0.
\label{eq:Sx_twoharm_2x2}
\end{equation}
Therefore the two-harmonic dispersion relation is
\begin{equation}
F_2(\omega,k)\equiv
\Big(A-\varepsilon_0\omega^2\Big)\Big(A-\varepsilon_0(\omega-\Omega)^2\Big)
-\omega^2(\omega-\Omega)^2\,\varepsilon_{+1}\varepsilon_{-1}=0,
\label{eq:Sx_twoharm_disp}
\end{equation}
where for real $\varepsilon(t)$ one has $\varepsilon_{-1}=\varepsilon_{+1}^*$ and $\varepsilon_{+1}\varepsilon_{-1}=|\varepsilon_1|^2\ge 0$. The geometrical band slope is obtained by implicit differentiation:
\begin{equation}
v_g(k)\equiv \frac{d\omega}{dk}
=
-\frac{\partial_kF_2}{\partial_\omega F_2},
\qquad F_2(\omega,k)=0.
\label{eq:Sx_twoharm_vg_implicit}
\end{equation}
Introducing $W\equiv \omega(\omega-\Omega)$ and $\Delta\equiv \varepsilon_0^2-\varepsilon_{+1}\varepsilon_{-1}$, one convenient on-shell form is
\begin{equation}
v_g(k)
=
\frac{k}{\mu}\,
\frac{2A-\varepsilon_0\big[\omega^2+(\omega-\Omega)^2\big]}
{(2\omega-\Omega)\big(\varepsilon_0A-\Delta W\big)},
\qquad A=\frac{k^2}{\mu},\quad F_2(\omega,k)=0.
\label{eq:Sx_twoharm_vg_closed}
\end{equation}

\noindent
For the same two-harmonic eigenmode, the energy-transport velocity $v_E=\langle S_z\rangle_T/\langle u\rangle_T$ can be evaluated using Eq.~\eqref{eq:Sx.vE_bilinear} together with $H_0=(k/\mu\omega)E_0$ and $H_{-1}=(k/\mu(\omega-\Omega))E_{-1}$, and eliminating the amplitude ratio with Eq.~\eqref{eq:Sx_twoharm_2x2}. After straightforward algebra this yields the compact on-shell expression
\begin{equation}
v_E(k)
=
\frac{(2\omega-\Omega)\big(k^2-\mu\varepsilon_0 W\big)}
{k\Big(2k^2-\mu\varepsilon_0\big[\omega^2+(\omega-\Omega)^2\big]\Big)},
\qquad W=\omega(\omega-\Omega),\quad F_2(\omega,k)=0.
\label{eq:Sx_twoharm_vE_closed}
\end{equation}

\noindent
Finally, the product simplifies to a particularly compact form:
\begin{equation}
v_g(k)\,v_E(k)
=
\frac{A-\varepsilon_0 W}{\mu\big(\varepsilon_0A-\Delta W\big)},
\qquad A=\frac{k^2}{\mu},\quad W=\omega(\omega-\Omega),\quad F_2(\omega,k)=0.
\label{eq:Sx_twoharm_product}
\end{equation}
Equation~\eqref{eq:Sx_twoharm_product} is useful as an analytic benchmark and makes explicit that near a symmetry-selected square-root edge, the divergence of $v_g$ is paired with the collapse of $v_E$ such that the product remains finite, consistent with Eq.~\eqref{eq:Sx_vgvE_finite}.

\section{\,\,\,\,\,\, Relation between group and energy-transport velocity}
\label{sec:SN4_vgvE}

\noindent In this Supplementary Note we derive an exact ``velocity-product law'' on any stable (real) Floquet band of a homogeneous, temporally periodic medium:
\begin{equation}
v_E\,v_g=\bigg\langle\frac{1}{\mu\varepsilon(t)}\bigg\rangle_T
=\big\langle v_{\mathrm{ph}}^2(t)\big\rangle_T,
\label{eq:SN4_vgvE_final}
\end{equation}
where $v_g=\partial_k\omega(k)$ is the group velocity, $v_E\equiv \braket{S_z}_T/\braket{u}_T$ is the cycle-averaged energy-transport velocity, and
\begin{equation}
v_{\mathrm{ph}}(t)\equiv \frac{1}{\sqrt{\mu\varepsilon(t)}}.
\end{equation}

\subsection{Equations of motion rearrangement and $\sigma_x$-pseudo Hermiticity}
\noindent Maxwell's equations for a 1D plane wave ($\propto e^{ikz}$) in a time-varying permittivity $\varepsilon(t)$ read
\begin{equation}
\begin{aligned}
&i\partial_t H(t)=k\mu^{-1}E(t),\\
&i\partial_t\!\big(\varepsilon(t)E(t)\big)=kH(t).
\end{aligned}
\label{eq:SN4_maxwell}
\end{equation}
For this Note we use the auxiliary basis
\begin{equation}
\mathbf\Psi(t)\equiv
\begin{pmatrix}\sqrt{\varepsilon(t)}\,E(t)\\ \sqrt{\mu}\,H(t)\end{pmatrix}.
\label{eq:SN4_basis}
\end{equation}
Rewriting Eq.~\eqref{eq:SN4_maxwell} gives
\begin{equation}
\begin{aligned}
&i\partial_t\!\big(\sqrt{\mu}\,H\big)=\frac{k}{\sqrt{\mu\varepsilon(t)}}\big(\sqrt{\varepsilon(t)}\,E\big),\\
&i\partial_t\!\big(\sqrt{\varepsilon(t)}\,E\big)=\frac{k}{\sqrt{\mu\varepsilon(t)}}\big(\sqrt{\mu}\,H\big)
-\frac{i\dot{\varepsilon}(t)}{2\varepsilon(t)}\big(\sqrt{\varepsilon(t)}\,E\big),
\end{aligned}
\label{eq:SN4_rewritten}
\end{equation}
and hence
\begin{equation}
\begin{aligned}
i\partial_t\mathbf\Psi(t)
&=
\begin{pmatrix}
-\dfrac{i\dot{\varepsilon}}{2\varepsilon} & \dfrac{k}{\sqrt{\varepsilon\mu}}\\[4pt]
\dfrac{k}{\sqrt{\varepsilon\mu}} & 0
\end{pmatrix}\mathbf\Psi(t)\\
&\equiv \bigg[-\dfrac{i\dot{\varepsilon}}{4\varepsilon}\,\mathbb{I}+\mathcal{H}(t)\bigg]\mathbf\Psi(t),
\end{aligned}
\label{eq:SN4_eom_matrix}
\end{equation}
with the $T$-periodic matrix
\begin{equation}
\mathcal{H}(t)=
\begin{bmatrix}
-\dfrac{i\dot{\varepsilon}}{4\varepsilon} & \dfrac{k}{\sqrt{\varepsilon \mu}}\\[4pt]
\dfrac{k}{\sqrt{\varepsilon \mu}} & \dfrac{i\dot{\varepsilon}}{4\varepsilon}
\end{bmatrix},
\qquad
\sigma_x\mathcal{H}(t)\sigma_x=\mathcal{H}^\dagger(t).
\label{eq:SN4_pseudoH}
\end{equation}

\noindent Let $\mathcal{U}(t_1,t_2)$ denote the evolution operator generated by Eq.~\eqref{eq:SN4_eom_matrix}.
From Eq.~\eqref{eq:SN4_pseudoH} one obtains the generalized $\sigma_x$-pseudo unitarity
\begin{equation}
\mathcal{U}^\dagger(t_1,t_2)\,\sigma_x\,\mathcal{U}(t_1,t_2)
=\sqrt{\frac{\varepsilon(t_2)}{\varepsilon(t_1)}}\,\sigma_x,
\label{eq:SN4_pseudounitary_general}
\end{equation}
while over one full period $\varepsilon(T)=\varepsilon(0)$ implies that the monodromy matrix is strictly $\sigma_x$-pseudo unitary:
\begin{equation}
\mathcal{U}^\dagger(T,0)\,\sigma_x\,\mathcal{U}(T,0)=\sigma_x.
\label{eq:SN4_pseudounitary_monodromy}
\end{equation}

\subsection{Time-averaged Poynting flux $\braket{S_z}_T$}
\noindent The (cycle-averaged) Poynting flux is
\begin{equation}
\begin{aligned}
\braket{S_z}_T
&=\frac{1}{T}\int_0^T S_z(t)\,dt
=\frac{1}{T}\int_0^T \frac{1}{4}\!\left[E(t)H^*(t)+H(t)E^*(t)\right]dt.
\end{aligned}
\label{eq:SN4_Savg_def}
\end{equation}
Using $\mathbf\Psi$ in Eq.~\eqref{eq:SN4_basis},
\begin{equation}
\mathbf\Psi^\dagger(t)\sigma_x\mathbf\Psi(t)
=\sqrt{\varepsilon(t)\mu}\,\Big[E(t)H^*(t)+H(t)E^*(t)\Big],
\label{eq:SN4_sigmax_identity}
\end{equation}
and for a Floquet initial condition $\ket{R}\equiv \mathbf\Psi(0)$ we have
\begin{equation}
\mathbf\Psi(t)=\mathcal{U}(t,0)\ket{R}.
\end{equation}
Then Eq.~\eqref{eq:SN4_pseudounitary_general} with $(t_1,t_2)=(t,0)$ yields
\begin{equation}
\mathbf\Psi^\dagger(t)\sigma_x\mathbf\Psi(t)
=\bra{R}\mathcal{U}^\dagger(t,0)\sigma_x\mathcal{U}(t,0)\ket{R}
=\sqrt{\frac{\varepsilon(0)}{\varepsilon(t)}}\,\bra{R}\sigma_x\ket{R}.
\label{eq:SN4_sigmax_conserved_scaled}
\end{equation}
Combining Eqs.~\eqref{eq:SN4_sigmax_identity} and \eqref{eq:SN4_sigmax_conserved_scaled} gives
\begin{equation}
E(t)H^*(t)+H(t)E^*(t)
=\frac{\varepsilon(0)}{\varepsilon(t)}\Big[E(0)H^*(0)+H(0)E^*(0)\Big],
\label{eq:SN4_flux_scaling}
\end{equation}
equivalently $\varepsilon(t)S_z(t)=\varepsilon(0)S_z(0)$.
Substituting Eq.~\eqref{eq:SN4_flux_scaling} into Eq.~\eqref{eq:SN4_Savg_def} gives
\begin{equation}
\braket{S_z}_T
=\bigg\langle\frac{\varepsilon(0)}{\varepsilon(t)}\bigg\rangle_T S_z(0)
=\frac{\sqrt{\varepsilon(0)}}{4\sqrt{\mu}}\bigg\langle\frac{1}{\varepsilon(t)}\bigg\rangle_T \bra{R}\sigma_x\ket{R}.
\label{eq:SN4_Savg_result}
\end{equation}

\subsection{Time-averaged energy density $\braket{u}_T$}
\noindent The instantaneous energy density is
\begin{equation}
u(t)=\frac{1}{4}\Big(\varepsilon(t)|E(t)|^2+\mu|H(t)|^2\Big)
=\frac{1}{4}\,\mathbf\Psi^\dagger(t)\mathbf\Psi(t).
\end{equation}
Hence, for a Floquet initial condition $\ket{R}$,
\begin{equation}
\braket{u}_T
=\frac{1}{T}\int_0^T u(t)\,dt
=\frac{1}{4T}\int_0^T \bra{R}\mathcal{U}^\dagger(t,0)\mathcal{U}(t,0)\ket{R}\,dt.
\label{eq:SN4_uavg_result}
\end{equation}

\subsection{Group velocity $v_g=\partial_k\omega(k)$}
\noindent On a stable band, the monodromy matrix has an eigenpair
\begin{equation}
\mathcal{U}(T,0)\ket{R_k}=e^{-i\omega(k)T}\ket{R_k},
\label{eq:SN4_floquet_eig}
\end{equation}
with real $\omega(k)$ (so $|e^{-i\omega T}|=1$). Differentiating Eq.~\eqref{eq:SN4_floquet_eig} with respect to $k$ and projecting with the corresponding left eigenvector $\bra{L_k}$ gives
\begin{equation}
v_g=\partial_k\omega(k)
=\frac{i}{T}e^{i\omega(k)T}\frac{\bra{L_k}\,\partial_k\mathcal{U}(T,0)\,\ket{R_k}}{\braket{L_k|R_k}}.
\label{eq:SN4_vg_general}
\end{equation}
Using the strict $\sigma_x$-pseudo unitarity of $\mathcal{U}(T,0)$ in Eq.~\eqref{eq:SN4_pseudounitary_monodromy}, one may choose
\begin{equation}
\bra{L_k}=\bra{R_k}\sigma_x,
\qquad (\text{stable band}),
\label{eq:SN4_left_right}
\end{equation}
so that Eq.~\eqref{eq:SN4_vg_general} becomes
\begin{equation}
v_g=\frac{i}{T}e^{i\omega(k)T}
\frac{\bra{R_k}\sigma_x\,\partial_k\mathcal{U}(T,0)\,\ket{R_k}}{\bra{R_k}\sigma_x\ket{R_k}}.
\label{eq:SN4_vg_sigmax}
\end{equation}

\noindent To evaluate $\partial_k\mathcal{U}(T,0)$, note that $\mathcal{U}(t,0)$ obeys
\begin{equation}
i\partial_t\,\mathcal{U}(t,0)=H(k,t)\,\mathcal{U}(t,0),
\qquad \mathcal{U}(0,0)=\mathbb{I}_2,
\label{eq:SN4_U_eom}
\end{equation}
with $H(k,t)=-\frac{i\dot{\varepsilon}}{4\varepsilon}\mathbb{I}+\mathcal{H}(t)$, so that
\begin{equation}
\partial_k H(k,t)=\frac{1}{\sqrt{\varepsilon(t)\mu}}\,\sigma_x.
\label{eq:SN4_dkH}
\end{equation}
Standard variation of parameters yields
\begin{equation}
\partial_k\mathcal{U}(T,0)
=-i\int_0^T \mathcal{U}(T,\tau)\,\big(\partial_k H(k,\tau)\big)\,\mathcal{U}(\tau,0)\,d\tau.
\label{eq:SN4_dkU}
\end{equation}
Multiplying Eq.~\eqref{eq:SN4_dkU} by $\sigma_x$ on the left and using Eq.~\eqref{eq:SN4_dkH} gives
\begin{equation}
\sigma_x\partial_k\mathcal{U}(T,0)
=-i\int_0^T \frac{1}{\sqrt{\varepsilon(\tau)\mu}}\,
\sigma_x\mathcal{U}(T,\tau)\sigma_x\,\mathcal{U}(\tau,0)\,d\tau.
\label{eq:SN4_sigmax_dkU_1}
\end{equation}
From Eq.~\eqref{eq:SN4_pseudounitary_general} with $(t_1,t_2)=(T,\tau)$,
\begin{equation}
\sigma_x\mathcal{U}(T,\tau)\sigma_x
=\sqrt{\frac{\varepsilon(\tau)}{\varepsilon(T)}}\,\mathcal{U}^\dagger(\tau,T),
\label{eq:SN4_sigmaxU_sigmax}
\end{equation}
and with $\varepsilon(T)=\varepsilon(0)$ one obtains
\begin{equation}
\sigma_x\partial_k\mathcal{U}(T,0)
=-i\int_0^T \frac{1}{\sqrt{\varepsilon(0)\mu}}\,
\mathcal{U}^\dagger(\tau,T)\,\mathcal{U}(\tau,0)\,d\tau.
\label{eq:SN4_sigmax_dkU_2}
\end{equation}

\noindent Substituting Eq.~\eqref{eq:SN4_sigmax_dkU_2} into Eq.~\eqref{eq:SN4_vg_sigmax} and using
\begin{equation}
\bra{R_k}\mathcal{U}^\dagger(\tau,T)
=e^{-i\omega(k)T}\bra{R_k}\mathcal{U}^\dagger(\tau,0),
\label{eq:SN4_phase_pullout}
\end{equation}
one finally finds
\begin{equation}
v_g
=\frac{1}{\bra{R_k}\sigma_x\ket{R_k}}
\bigg[\frac{1}{T}\int_0^T \frac{1}{\sqrt{\varepsilon(0)\mu}}\,
\bra{R_k}\mathcal{U}^\dagger(\tau,0)\mathcal{U}(\tau,0)\ket{R_k}\,d\tau\bigg].
\label{eq:SN4_vg_finalform}
\end{equation}

\subsection{Relation between $v_g$ and $v_E$}
\noindent By definition,
\begin{equation}
v_E\equiv\frac{\braket{S_z}_T}{\braket{u}_T}.
\label{eq:SN4_vE_def}
\end{equation}
Multiplying Eqs.~\eqref{eq:SN4_Savg_result}, \eqref{eq:SN4_uavg_result}, and \eqref{eq:SN4_vg_finalform} gives the exact identity
\begin{equation}
\begin{aligned}
v_E\,v_g
&=\bigg\langle\frac{1}{\mu\varepsilon(t)}\bigg\rangle_T
=\big\langle v_{\mathrm{ph}}^2(t)\big\rangle_T,
\end{aligned}
\label{eq:SN4_vgvE}
\end{equation}
which is Eq.~\eqref{eq:SN4_vgvE_final}.

\subsection{Equivalent Riemann-field representation}
\label{sec:SN4_Riemann_Phi}

\noindent The derivation above used the auxiliary field vector \(\mathbf\Psi=(\sqrt{\varepsilon}E,\sqrt{\mu}H)^T\), for which the Maxwell flux form is represented by \(\sigma_x\). The same structure can be written in a diagonal flux-form representation by introducing the Riemann field vector
\begin{equation}
\boldsymbol{\Phi}(t)
\equiv
\begin{pmatrix}
\Phi_+(t)\\
\Phi_-(t)
\end{pmatrix}
=
g(t)A\mathbf\Psi(t)
=
\varepsilon^{1/4}(t)
\begin{pmatrix}
\sqrt{\varepsilon(t)}\,E(t)+\sqrt{\mu}\,H(t)\\[2pt]
-\sqrt{\varepsilon(t)}\,E(t)+\sqrt{\mu}\,H(t)
\end{pmatrix},
\label{eq:SN4_Phi_basis_def}
\end{equation}
where
\begin{equation}
g(t)\equiv \varepsilon^{1/4}(t),
\qquad
A\equiv
\begin{pmatrix}
1 & 1\\
-1 & 1
\end{pmatrix}.
\label{eq:SN4_Phi_gA_def}
\end{equation}

With this convention, the diagonal quadratic form \(\boldsymbol{\Phi}^\dagger\sigma_z\boldsymbol{\Phi}=|\Phi_+|^2-|\Phi_-|^2\) is proportional to the Poynting flux. For constant \(\varepsilon\), \(\Phi_+\) and \(\Phi_-\) reduce, up to normalization and an overall sign, to the right- and left-going characteristic field combinations.

Using
\begin{equation}
A\sigma_xA^{-1}=\sigma_z,
\qquad
A\sigma_zA^{-1}=-\sigma_x,
\qquad
\frac{\dot g}{g}
=
\frac{1}{4}\frac{\dot{\varepsilon}}{\varepsilon},
\label{eq:SN4_A_identities}
\end{equation}
together with Eq.~\eqref{eq:SN4_eom_matrix}, one obtains
\begin{equation}
i\partial_t\boldsymbol{\Phi}(t)
=
\mathcal{H}_{\Phi}(k,t)\boldsymbol{\Phi}(t),
\label{eq:SN4_Phi_eom}
\end{equation}
where the time-evolution generator is
\begin{equation}
\mathcal{H}_{\Phi}(k,t)
=
k v_{\mathrm{ph}}(t)\sigma_z
+
i m(t)\sigma_x,
\qquad
m(t)\equiv
\frac{\dot{\varepsilon}(t)}{4\varepsilon(t)},
\qquad
v_{\mathrm{ph}}(t)
=
\frac{1}{\sqrt{\mu\varepsilon(t)}}.
\label{eq:SN4_Phi_generator}
\end{equation}
This equation is simply a change of field variables in the same Maxwell evolution. The scalar factor \(g(t)=\varepsilon^{1/4}(t)\) cancels the scalar trace term in Eq.~\eqref{eq:SN4_eom_matrix}, while the matrix
\(A\) maps the off-diagonal \(\sigma_x\) flux form to the diagonal \(\sigma_z\) form. Equation~\eqref{eq:SN4_Phi_generator} has the matrix form of a one-dimensional Dirac-type generator with an imaginary masslike coefficient \(i m(t)\).

For real \(k\) and real positive \(\varepsilon(t)\), the generator satisfies the flux-form relation
\begin{equation}
\mathcal{H}_{\Phi}^\dagger(k,t)\sigma_z
=
\sigma_z\mathcal{H}_{\Phi}(k,t).
\label{eq:SN4_Phi_flux_relation}
\end{equation}
Therefore, if \(\mathcal{U}_{\Phi}(t,0)\) denotes the propagator generated by Eq.~\eqref{eq:SN4_Phi_eom}, then it preserves the \(\sigma_z\) flux form:
\begin{equation}
\mathcal{U}_{\Phi}^\dagger(t,0)\sigma_z
\mathcal{U}_{\Phi}(t,0)
=
\sigma_z.
\label{eq:SN4_Phi_flux_conservation}
\end{equation}
Indeed,
\begin{equation}
\frac{d}{dt}
\left[
\mathcal{U}_{\Phi}^\dagger(t,0)\sigma_z
\mathcal{U}_{\Phi}(t,0)
\right]
=
i\mathcal{U}_{\Phi}^\dagger
\mathcal{H}_{\Phi}^\dagger\sigma_z
\mathcal{U}_{\Phi}
-
i\mathcal{U}_{\Phi}^\dagger
\sigma_z\mathcal{H}_{\Phi}
\mathcal{U}_{\Phi}
=
0,
\end{equation}
and \(\mathcal{U}_{\Phi}(0,0)=\mathbb{I}\).
For a solution
\begin{equation}
\boldsymbol{\Phi}(t)
=
\mathcal{U}_{\Phi}(t,0)\boldsymbol{\Phi}_0,
\label{eq:SN4_Phi_solution}
\end{equation}
this gives
\begin{equation}
\boldsymbol{\Phi}^\dagger(t)\sigma_z\boldsymbol{\Phi}(t)
=
\boldsymbol{\Phi}_0^\dagger\sigma_z\boldsymbol{\Phi}_0.
\label{eq:SN4_Phi_flux_constant}
\end{equation}
Thus the Riemann-field representation isolates a conserved diagonal quadratic form. Below we show that this form gives the Poynting flux, whereas the positive quadratic form \(\boldsymbol{\Phi}^\dagger\boldsymbol{\Phi}\) gives the electromagnetic energy density.

The physical energy density and Poynting flux take a simple form in this basis. From the inverse relations
\begin{equation}
\sqrt{\varepsilon(t)}\,E(t)
=
\frac{\Phi_+(t)-\Phi_-(t)}{2g(t)},
\qquad
\sqrt{\mu}\,H(t)
=
\frac{\Phi_+(t)+\Phi_-(t)}{2g(t)},
\label{eq:SN4_Phi_inverse}
\end{equation}
one finds
\begin{equation}
u(t)
=
\frac{1}{8g^2(t)}
\boldsymbol{\Phi}^\dagger(t)\boldsymbol{\Phi}(t),
\label{eq:SN4_Phi_energy_density}
\end{equation}
and
\begin{equation}
S_z(t)
=
\frac{v_{\mathrm{ph}}(t)}{8g^2(t)}
\boldsymbol{\Phi}^\dagger(t)\sigma_z\boldsymbol{\Phi}(t).
\label{eq:SN4_Phi_flux_density}
\end{equation}
Thus the positive form \(\boldsymbol{\Phi}^\dagger\boldsymbol{\Phi}\) determines the energy density, while the indefinite diagonal form \(\boldsymbol{\Phi}^\dagger\sigma_z\boldsymbol{\Phi}\) determines the Poynting flux.

For a Floquet solution, Eq.~\eqref{eq:SN4_Phi_flux_constant} gives
\begin{equation}
\braket{S_z}_T
=
\bar a_{\Phi}\,
\boldsymbol{\Phi}_0^\dagger\sigma_z\boldsymbol{\Phi}_0,
\label{eq:SN4_Phi_Savg}
\end{equation}
with
\begin{equation}
\bar a_{\Phi}
\equiv
\bigg\langle
\frac{v_{\mathrm{ph}}(t)}{8g^2(t)}
\bigg\rangle_T.
\label{eq:SN4_Phi_a_bar}
\end{equation}

The energy denominator can be written using a positive period-averaged Gramian:
\begin{equation}
\braket{u}_T
=
\boldsymbol{\Phi}_0^\dagger
G_u(k)
\boldsymbol{\Phi}_0,
\label{eq:SN4_Phi_uavg_Gramian}
\end{equation}
where
\begin{equation}
G_u(k)
=
\frac{1}{T}
\int_0^T
\mathcal{U}_{\Phi}^\dagger(t,0)
\frac{\mathbb{I}}{8g^2(t)}
\mathcal{U}_{\Phi}(t,0)\,dt .
\label{eq:SN4_Phi_Gu_def}
\end{equation}
Here \(G_u(k)\) is the period-averaged matrix associated with the energy quadratic form in the Riemann-field representation; the quadratic form \(\boldsymbol{\Phi}_0^\dagger G_u(k)\boldsymbol{\Phi}_0\) gives the cycle-averaged energy density \(\braket{u}_T\).

To see that \(G_u(k)\) is positive definite, take any nonzero vector
\(\mathbf{x}\). Then
\begin{equation}
\begin{aligned}
\mathbf{x}^\dagger G_u(k)\mathbf{x}
&=
\frac{1}{T}
\int_0^T
\mathbf{x}^\dagger
\mathcal{U}_{\Phi}^\dagger(t,0)
\frac{\mathbb{I}}{8g^2(t)}
\mathcal{U}_{\Phi}(t,0)
\mathbf{x}\,dt  \\
&=
\frac{1}{T}
\int_0^T
\frac{
\left\|\mathcal{U}_{\Phi}(t,0)\mathbf{x}\right\|^2
}{8g^2(t)}\,dt .
\end{aligned}
\label{eq:SN4_Phi_Gu_positive_proof}
\end{equation}
Since \(\varepsilon(t)>0\), we have \(g^2(t)>0\). Moreover,
\(\mathcal{U}_{\Phi}(t,0)\) is invertible for finite \(t\), so
\(\mathcal{U}_{\Phi}(t,0)\mathbf{x}\neq0\) for any
\(\mathbf{x}\neq0\). Therefore
\begin{equation}
\mathbf{x}^\dagger G_u(k)\mathbf{x}>0
\qquad
\text{for all } \mathbf{x}\neq0,
\end{equation}
or equivalently
\begin{equation}
G_u(k)\succ0 .
\label{eq:SN4_Phi_Gu_positive}
\end{equation}

Consequently, the energy-transport velocity may be written as
\begin{equation}
v_E(k)
=
\frac{\braket{S_z}_T}{\braket{u}_T}
=
\bar a_{\Phi}
\frac{
\boldsymbol{\Phi}_0^\dagger\sigma_z\boldsymbol{\Phi}_0
}{
\boldsymbol{\Phi}_0^\dagger G_u(k)\boldsymbol{\Phi}_0
}.
\label{eq:SN4_Phi_vE_fluxform}
\end{equation}
This equivalent Riemann-field representation makes explicit that, for a finite nonzero normalization of the field vector, the gap-edge vanishing of \(v_E\) corresponds to a collapse of the Poynting-flux numerator,
\begin{equation}
\boldsymbol{\Phi}_0^\dagger\sigma_z\boldsymbol{\Phi}_0\to0,
\label{eq:SN4_Phi_flux_null}
\end{equation}
while the cycle-averaged energy denominator remains positive and finite. Thus the gap-edge suppression of the energy-transport velocity is a classical flux-form effect, not a divergence or singularity of the cycle-averaged electromagnetic energy density.

\section{\,\,\,\,\,\, Energy centroid and modulation-driven term}
\label{sec:centroid}

\noindent In the main text we use the energy centroid to relate band velocities to the motion of a wavepacket.
Here we provide the detailed derivation of the centroid equation and show how the modulation term can be expressed in terms of global electric and magnetic centroids.

\subsection{Derivation of the centroid equation}

\noindent We define the total energy and centroid as
\begin{equation}
U(t) = \int_{-\infty}^{\infty} u(z,t)\,dz,
\qquad
Z(t) = \frac{\int_{-\infty}^{\infty} z\,u(z,t)\,dz}{U(t)}.
\label{eq:S3.Z_def}
\end{equation}
We assume a localized, finite-energy wavepacket such that $0<U(t)<\infty$, $\int_{-\infty}^{\infty}|z|\,u(z,t)\,dz<\infty$, and
$|z|\,|S_z(z,t)|\to 0$ as $|z|\to\infty$, so that boundary terms vanish below.
Differentiating and using the quotient rule, we obtain
\begin{equation}
\dot Z(t)
= \frac{\int z\,\partial_t u\,dz}{U(t)}
 - \frac{\int z\,u\,dz}{U^2(t)}\,\partial_t U(t)
= \frac{1}{U(t)}\int (z-Z)\,\partial_t u(z,t)\,dz.
\label{eq:S3.Zdot_step}
\end{equation}
Using the Poynting theorem
\begin{equation}
\partial_t u + \partial_z S_z = -\tfrac14\dot\varepsilon(t)|E|^2,
\end{equation}
we can write
\begin{equation}
\partial_t u = -\partial_z S_z - \tfrac14\dot\varepsilon(t)|E|^2.
\end{equation}
Substituting this into Eq.~\eqref{eq:S3.Zdot_step} and integrating by parts, we obtain
\begin{align}
\int (z-Z)\,\partial_t u\,dz
&= -\int (z-Z)\,\partial_z S_z\,dz
    -\frac14\dot\varepsilon(t)\int (z-Z)|E|^2\,dz\notag\\
&= -\bigl[(z-Z)S_z\bigr]_{-\infty}^{\infty}
   +\int S_z\,dz
   -\frac14\dot\varepsilon(t)\int (z-Z)|E|^2\,dz.
\end{align}
By the localization assumption, the boundary term vanishes and we obtain
\begin{equation}
\int (z-Z)\,\partial_t u\,dz
= \int S_z(z,t)\,dz
 -\frac14\dot\varepsilon(t)\int (z-Z)|E(z,t)|^2\,dz.
\end{equation}
Inserting this into Eq.~\eqref{eq:S3.Zdot_step} yields
\begin{equation}
\dot Z(t)
= \frac{\int S_z(z,t)\,dz}{U(t)}
 - \frac{1}{4U(t)}\int (z-Z(t))\,\dot\varepsilon(t)\,|E(z,t)|^2\,dz.
\label{eq:S3.Zdot_general}
\end{equation}

\subsection{Modulation term in terms of electric and magnetic centroids}

\noindent To make the modulation term more transparent, we introduce the electric and magnetic intensity moments
\begin{equation}
I_E(t) = \int_{-\infty}^{\infty} |E(z,t)|^2\,dz,\qquad
I_H(t) = \int_{-\infty}^{\infty} |H(z,t)|^2\,dz,
\end{equation}
and their centroids
\begin{equation}
Z_E(t) = \frac{\int_{-\infty}^{\infty} z\,|E|^2 dz}{I_E},\qquad
Z_H(t) = \frac{\int_{-\infty}^{\infty} z\,|H|^2 dz}{I_H}.
\end{equation}
In a spatially uniform medium these coincide with the corresponding electric and magnetic \emph{energy} centroids, since $\varepsilon(t)$ and $\mu$ are $z$-independent. We also define the electric and magnetic energies
\begin{equation}
U_E(t) = \frac14 \varepsilon(t) I_E(t),\qquad
U_H(t) = \frac14 \mu I_H(t),
\end{equation}
so that \(U(t)=U_E(t)+U_H(t)\).
With these definitions we can express the global centroid as
\begin{equation}
Z(t) = \frac{U_E Z_E + U_H Z_H}{U_E+U_H}
\equiv f_E(t)\, Z_E(t) + f_H(t)\, Z_H(t),
\end{equation}
where
\begin{equation}
f_E(t) \equiv \frac{U_E}{U},\qquad
f_H(t) \equiv \frac{U_H}{U},
\end{equation}
are the electric and magnetic energy fractions, with \(f_E+f_H=1\). The numerator of the modulation term in Eq.~\eqref{eq:S3.Zdot_general} can be written as
\begin{equation}
\int (z-Z)|E|^2 dz
= \int z|E|^2 dz - Z \int |E|^2 dz
= I_E Z_E - Z I_E
= I_E\bigl(Z_E - Z\bigr).
\end{equation}
Using the relation \(Z=f_E Z_E+f_H Z_H\), we find
\begin{equation}
Z_E - Z
= Z_E - (f_E Z_E + f_H Z_H)
= f_H\bigl(Z_E - Z_H\bigr).
\end{equation}
Thus
\begin{equation}
\int (z-Z)|E|^2 dz
= I_E f_H\bigl(Z_E - Z_H\bigr).
\end{equation}
Substituting into Eq.~\eqref{eq:S3.Zdot_general}, we obtain
\begin{equation}
-\frac{1}{4U}\int (z-Z)\dot\varepsilon |E|^2 dz
= -\frac{\dot\varepsilon(t)}{4U(t)}\,I_E(t) f_H(t)\bigl[Z_E(t)-Z_H(t)\bigr].
\end{equation}
Using \(U_E=\tfrac14\varepsilon I_E\) and \(f_E=U_E/U\), i.e.\ \(I_E = 4U_E/\varepsilon = 4f_E U/\varepsilon\), we can rewrite this as
\begin{equation}
-\frac{1}{4U}\int (z-Z)\dot\varepsilon |E|^2 dz
= -\frac{\dot\varepsilon(t)}{\varepsilon(t)}\,f_E(t)f_H(t)\bigl[Z_E(t)-Z_H(t)\bigr].
\label{eq:S3.mod_term_final}
\end{equation}
In other words, the centroid equation can be written as
\begin{equation}
\dot Z(t)
= v_{\rm flux}(t) + v_{\rm mod}(t),
\end{equation}
with
\begin{equation}
v_{\rm flux}(t) \equiv \frac{\int_{-\infty}^{\infty} S_z(z,t)\,dz}{U(t)},
\qquad
v_{\rm mod}(t) \equiv -\frac{\dot\varepsilon(t)}{\varepsilon(t)}\,f_E(t)f_H(t)\bigl[Z_E(t)-Z_H(t)\bigr].
\end{equation}
The flux term $v_{\rm flux}(t)$ is the instantaneous transport contribution (total flux divided by total energy). The modulation term is controlled by three factors: the relative modulation rate \(\dot\varepsilon/\varepsilon\), the simultaneous presence of electric and magnetic energy (through the product \(f_E f_H\)), and the separation between the electric and magnetic centroids \(Z_E-Z_H\). In smooth passbands, where \(|E|^2\) and \(|H|^2\) have nearly identical spatial profiles and \(Z_E\approx Z_H\), the modulation term is parametrically small and the centroid velocity is well approximated by the band energy velocity.
Near band edges and exceptional points, forward and backward components mix into standing-wave patterns, the electric and magnetic energy distributions become strongly out of phase (so $|Z_E-Z_H|$ grows), and the modulation term can become comparable to the flux term; in that regime the centroid motion reflects a nontrivial interplay between intrinsic energy transport and modulation-driven energy exchange with the temporal drive.

\section{\,\,\,\,\,\, Floquet Wavepackets in the PTC}
\label{sec:wp_vflux_vmod_residual}

\noindent
In this Note we analyze energy-centroid dynamics for \emph{passband} wavepackets in a spatially uniform PTC.
Because the medium is homogeneous in space, each component carries a plane-wave factor $e^{ikz}$. We construct wavepackets by superposing Floquet plane-wave modes from a \emph{single} real branch $\omega(k)\in\mathbb R$,
with $k$-support chosen to remain strictly outside the momentum gap. Using the centroid identity from~\ref{sec:centroid}, $\dot Z(t)=v_{\rm flux}(t)+v_{\rm mod}(t)$, we derive exact $k$-space formulas for $v_{\rm flux}(t)$ and express $v_{\rm mod}(t)$ in terms of global electric/magnetic centroids, together with narrowband period-averaged limits.

\subsection{Passband Floquet eigenmodes and wavepacket construction}
\label{subsec:wp_definition}

\noindent
For each passband $k$, the truncated Floquet plane-wave mode can be written as
\begin{equation}
E_k(z,t)=e^{ikz-i\omega(k)t}\,\mathcal E_k(t),
\qquad
H_k(z,t)=e^{ikz-i\omega(k)t}\,\mathcal H_k(t),
\label{eq:SI_mode_periodic_parts}
\end{equation}
where the periodic parts are 
\begin{equation}
\mathcal E_k(t)\equiv \sum_mE_m(k)e^{-im\Omega t},
\qquad
\mathcal H_k(t)\equiv \sum_mH_m(k)e^{-im\Omega t}.
\label{eq:SI_periodic_parts_2harm}
\end{equation}
For a wavepacket we superpose passband components of a \emph{single} real Floquet branch $\omega(k)\in\mathbb R$, with a continuously chosen eigenvector gauge in $k$:
\begin{equation}
E(z,t)=\int_{\mathbb R}\frac{dk}{2\pi}\,A(k)\,\mathcal E_k(t)\,e^{ikz-i\omega(k)t},
\qquad
H(z,t)=\int_{\mathbb R}\frac{dk}{2\pi}\,A(k)\,\mathcal H_k(t)\,e^{ikz-i\omega(k)t},
\label{eq:SI_wp_fields}
\end{equation}
where $A(k)$ is a localized envelope whose effective support is contained in the passband.
To ensure that no momentum-gap components enter, we choose $k_0$ in the passband and a bandwidth $\sigma_k$ such that
$A(k)$ is negligible whenever $\omega(k)$ becomes complex.

\medskip
\noindent
Let $\tilde E(k,t)$ and $\tilde H(k,t)$ denote the spatial Fourier transforms consistent with
\begin{equation}
E(z,t)=\int\frac{dk}{2\pi}\,\tilde E(k,t)e^{ikz},
\qquad
H(z,t)=\int\frac{dk}{2\pi}\,\tilde H(k,t)e^{ikz}.
\end{equation}
For the wavepacket \eqref{eq:SI_wp_fields},
\begin{equation}
\tilde E(k,t)=A(k)\,\mathcal E_k(t)\,e^{-i\omega(k)t},\qquad
\tilde H(k,t)=A(k)\,\mathcal H_k(t)\,e^{-i\omega(k)t}.
\label{eq:SI_tildeEH}
\end{equation}

\subsection{$v_{\rm flux}(t)$ for the wavepacket and its period average}
\label{subsec:vflux_wavepacket}

\noindent
The instantaneous total energy and total flux are
\begin{equation}
U(t)=\int u(z,t)\,dz
=\frac14\varepsilon(t)\int |E|^2dz+\frac14\mu\int |H|^2dz,
\qquad
\mathcal S(t)=\int S_z(z,t)\,dz=\frac12\Re\int E H^*dz,
\end{equation}
and the flux velocity is $v_{\rm flux}(t)=\mathcal S(t)/U(t)$.
Using Parseval,
\begin{equation}
\int |E|^2dz=\int\frac{dk}{2\pi}\,|\tilde E(k,t)|^2,
\qquad
\int |H|^2dz=\int\frac{dk}{2\pi}\,|\tilde H(k,t)|^2,
\qquad
\int E H^*dz=\int\frac{dk}{2\pi}\,\tilde E(k,t)\tilde H^*(k,t),
\end{equation}
and using \eqref{eq:SI_tildeEH} (the overall phase $e^{-i\omega t}$ cancels in quadratic quantities), we obtain the \emph{exact} $k$-space expressions
\begin{align}
U(t)
&=\int\frac{dk}{2\pi}\,|A(k)|^2\,u_k(t),
\qquad
u_k(t)\equiv \frac14\Big(\varepsilon(t)\,|\mathcal E_k(t)|^2+\mu\,|\mathcal H_k(t)|^2\Big),
\label{eq:SI_Uk_def}\\[2pt]
\mathcal S(t)
&=\int\frac{dk}{2\pi}\,|A(k)|^2\,s_k(t),
\qquad
s_k(t)\equiv \frac12\Re\big(\mathcal E_k(t)\,\mathcal H_k^*(t)\big).
\label{eq:SI_Phik_def}
\end{align}
Therefore
\begin{equation}
v_{\rm flux}(t)
=
\frac{\displaystyle \int\frac{dk}{2\pi}\,|A(k)|^2\,s_k(t)}
     {\displaystyle \int\frac{dk}{2\pi}\,|A(k)|^2\,u_k(t)}.
\label{eq:SI_vflux_exact}
\end{equation}
We define the period average over one modulation cycle as
\begin{equation}
\langle v_{\rm flux}\rangle_T\equiv \frac{1}{T}\int_0^T v_{\rm flux}(t)\,dt.
\label{eq:SI_vflux_timeavg_def}
\end{equation}

\medskip
\noindent
Let $A(k)$ be real and localized about $k_0$ with characteristic width $\sigma_k$, and assume the \emph{centering condition}
\begin{equation}
\int_{\mathbb R}\frac{dk}{2\pi}\,(k-k_0)\,|A(k)|^2=0
\qquad
\text{(e.g., $|A(k)|^2$ is even about $k_0$).}
\label{eq:SI_centering_condition}
\end{equation}
If $s_k(t)$ and $u_k(t)$ are sufficiently smooth in $k$ across the effective support of $A(k)$, then
\begin{equation}
v_{\rm flux}(t)=\frac{s_{k_0}(t)}{u_{k_0}(t)}+O(\sigma_k^2),
\qquad
\langle v_{\rm flux}\rangle_T
=
\frac{1}{T}\int_0^T\frac{s_{k_0}(t)}{u_{k_0}(t)}\,dt
+O(\sigma_k^2).
\label{eq:SI_vflux_narrowband}
\end{equation}
Without the centering condition \eqref{eq:SI_centering_condition}, the leading correction is generically $O(\sigma_k)$. For comparison, the \emph{band energy velocity} used in the main text is
\begin{equation}
v_E(k_0)=\frac{\langle s_{k_0}\rangle_T}{\langle u_{k_0}\rangle_T}
\qquad
(\langle f\rangle_T\equiv T^{-1}\int_0^T f(t)\,dt),
\label{eq:SI_vE_vs_vfluxavg}
\end{equation}
which in general differs from the time average of the ratio in Eq.~\eqref{eq:SI_vflux_narrowband} unless $s_{k_0}(t)/u_{k_0}(t)$ is time-independent.

\subsection{$v_{\rm mod}(t)$ for the wavepacket and its period average}
\label{subsec:vmod_wavepacket}

\noindent
The energy-centroid decomposition derived in~\ref{sec:centroid} reads
\begin{equation}
\dot Z(t)=v_{\rm flux}(t)+v_{\rm mod}(t),
\qquad
v_{\rm mod}(t)
=
-\frac{\dot\varepsilon(t)}{\varepsilon(t)}\,f_E(t)f_H(t)\,\big[Z_E(t)-Z_H(t)\big],
\label{eq:SI_Zdot_decomp_again}
\end{equation}
with $f_{E,H}(t)=U_{E,H}(t)/U(t)$ and
\begin{equation}
Z_E(t)=\frac{\int z|E|^2dz}{\int |E|^2dz},
\qquad
Z_H(t)=\frac{\int z|H|^2dz}{\int |H|^2dz}.
\end{equation}
To express $Z_E$ and $Z_H$ in $k$-space we use the Fourier moment identity (valid for localized wavepackets with finite first moment)~\cite{SM-SundaramNiu1999WavepacketDynamics}
\begin{equation}
\int z\,|E(z,t)|^2dz
=\int\frac{dk}{2\pi}\,\tilde E^*(k,t)\,\big(i\partial_k\big)\tilde E(k,t),
\qquad
\int z\,|H(z,t)|^2dz
=\int\frac{dk}{2\pi}\,\tilde H^*(k,t)\,\big(i\partial_k\big)\tilde H(k,t).
\label{eq:SI_moment_identity}
\end{equation}
Inserting \eqref{eq:SI_tildeEH} and assuming a real envelope $A(k)$ (with vanishing boundary terms in $k$-space) yields
\begin{align}
Z_E(t)
&=
t\,\frac{\displaystyle \int\frac{dk}{2\pi}\,|A(k)|^2\,|\mathcal E_k(t)|^2\,v_g(k)}
        {\displaystyle \int\frac{dk}{2\pi}\,|A(k)|^2\,|\mathcal E_k(t)|^2}
-
\frac{\displaystyle \int\frac{dk}{2\pi}\,|A(k)|^2\,\Im\!\big(\mathcal E_k^*(t)\,\partial_k\mathcal E_k(t)\big)}
     {\displaystyle \int\frac{dk}{2\pi}\,|A(k)|^2\,|\mathcal E_k(t)|^2},
\label{eq:SI_ZE_exact}\\[3pt]
Z_H(t)
&=
t\,\frac{\displaystyle \int\frac{dk}{2\pi}\,|A(k)|^2\,|\mathcal H_k(t)|^2\,v_g(k)}
        {\displaystyle \int\frac{dk}{2\pi}\,|A(k)|^2\,|\mathcal H_k(t)|^2}
-
\frac{\displaystyle \int\frac{dk}{2\pi}\,|A(k)|^2\,\Im\!\big(\mathcal H_k^*(t)\,\partial_k\mathcal H_k(t)\big)}
     {\displaystyle \int\frac{dk}{2\pi}\,|A(k)|^2\,|\mathcal H_k(t)|^2}.
\label{eq:SI_ZH_exact}
\end{align}
Here, the total-derivative term generated by $i\partial_k$ is purely imaginary and therefore drops out of the real moment. Assume in addition the centering condition~\eqref{eq:SI_centering_condition} and smoothness in $k$ across the effective support of $A(k)$.
Then, to leading order in the bandwidth $\sigma_k$,
\begin{equation}
Z_E(t)=t\,v_g(k_0)+\mathcal A_E(k_0,t)+O(\sigma_k^2),
\qquad
Z_H(t)=t\,v_g(k_0)+\mathcal A_H(k_0,t)+O(\sigma_k^2),
\label{eq:SI_ZE_ZH_narrowband}
\end{equation}
where we introduced the geometric k-space phase connections of the electric/magnetic Floquet envelopes
\begin{equation}
\mathcal A_E(k,t)\equiv
-\frac{\Im\!\big(\mathcal E_k^*(t)\,\partial_k\mathcal E_k(t)\big)}{|\mathcal E_k(t)|^2},
\qquad
\mathcal A_H(k,t)\equiv
-\frac{\Im\!\big(\mathcal H_k^*(t)\,\partial_k\mathcal H_k(t)\big)}{|\mathcal H_k(t)|^2}.
\label{eq:SI_connections_def}
\end{equation}
Therefore,
\begin{equation}
Z_E(t)-Z_H(t)
=
\Delta\mathcal A(k_0,t)+O(\sigma_k^2),
\qquad
\Delta\mathcal A(k,t)\equiv \mathcal A_E(k,t)-\mathcal A_H(k,t).
\label{eq:SI_ZEminusZH_connection}
\end{equation}
Under a common rephasing \(\mathcal E_k\!\to e^{i\chi(k)}\mathcal E_k\), \(\mathcal H_k\!\to e^{i\chi(k)}\mathcal H_k\),
both \(\mathcal A_E\) and \(\mathcal A_H\) shift by \( -\partial_k\chi\), but their difference
\begin{equation}
\Delta\mathcal A(k,t)\equiv \mathcal A_E(k,t)-\mathcal A_H(k,t)
\end{equation}
is gauge invariant. Substituting Eq.~\eqref{eq:SI_ZEminusZH_connection} into Eq.~\eqref{eq:SI_Zdot_decomp_again} yields
\begin{equation}
v_{\rm mod}(t)
=
-\frac{\dot\varepsilon(t)}{\varepsilon(t)}\,f_E(t)f_H(t)\,\Delta\mathcal A(k_0,t)
+O(\sigma_k^2).
\label{eq:SI_vmod_connection}
\end{equation}
In the same narrowband regime, the energy fractions may be approximated by the mode-resolved intensities,
\begin{equation}
f_E(t)
=
\frac{\varepsilon(t)\,|\mathcal E_{k_0}(t)|^2}{\varepsilon(t)\,|\mathcal E_{k_0}(t)|^2+\mu\,|\mathcal H_{k_0}(t)|^2}
+O(\sigma_k^2),
\qquad
f_H(t)=1-f_E(t).
\label{eq:SI_fE_fH_narrowband}
\end{equation}
We define the period average
\begin{equation}
\langle v_{\rm mod}\rangle_T\equiv \frac{1}{T}\int_0^T v_{\rm mod}(t)\,dt.
\label{eq:SI_vmod_timeavg_def}
\end{equation}
Since \(\frac{1}{T}\int_0^T (\dot\varepsilon/\varepsilon)\,dt=0\) by periodicity, a nonzero \(\langle v_{\rm mod}\rangle_T\) arises from the
\emph{correlation} between \(\dot\varepsilon/\varepsilon\) and the time-dependent factor \(f_E f_H\,\Delta\mathcal A\).

\subsection{Physical interpretation of the geometric \(k\)-space phase connections}
\label{subsec:SI_phase_connections_interp}

\noindent
Equations~\eqref{eq:SI_ZE_ZH_narrowband}--\eqref{eq:SI_connections_def} show that \(i\partial_k\) acts as the generator of first moments in real space,
so the electric and magnetic centroids contain, in addition to the ballistic term \(t\,v_g\), offsets controlled by the \(k\)-dependence of the Floquet envelopes.
Writing \(\mathcal E_k(t)=|\mathcal E_k(t)|e^{i\theta_E(k,t)}\) and \(\mathcal H_k(t)=|\mathcal H_k(t)|e^{i\theta_H(k,t)}\),
one has \(\mathcal A_E=-\partial_k\theta_E\) and \(\mathcal A_H=-\partial_k\theta_H\), i.e.geometric \(k\)-space phase connections.
Their gauge-invariant difference \(\Delta\mathcal A\) directly quantifies the instantaneous electric--magnetic centroid separation \(Z_E-Z_H\)
[Eq.~\eqref{eq:SI_ZEminusZH_connection}], and therefore controls the modulation-driven centroid drift through Eq.~\eqref{eq:SI_vmod_connection}.
Physically, temporal modulation reweights the electric and magnetic energy fractions; if their centroids are separated, this reweighting shifts the total center of energy
even when the net Poynting-flux transport remains bounded.

\section{\,\,\,\,\,\, Extension to dispersive and weakly lossy media: driven Lorentz model}
\label{sec:lorentz}

\noindent
The main-text analysis assumes a nondispersive, lossless permittivity \(\varepsilon(t)\).
In realistic platforms the permittivity is frequency-dispersive and may exhibit weak intrinsic loss.
In this Note we outline how the flux-over-energy interpretation extends to such media using a driven Lorentz model similar in spirit to that employed in prior work on photonic time crystals~\cite{SM-park2025spontaneous}.

\subsection{Maxwell--Lorentz model and total energy balance}
\label{subsec:lorentz_model}

\noindent
We consider a spatially uniform medium whose polarization \(P(z,t)\) obeys a driven Lorentz equation with a time-periodic stiffness~\cite{SM-LandauLifshitzECM,SM-Jackson1998ClassicalElectrodynamics,SM-park2025spontaneous},
\begin{equation}
\ddot P(z,t) + \eta\,\dot P(z,t) + \kappa(t)\,P(z,t)
= \alpha\,E(z,t),
\qquad
\kappa(t+T)=\kappa(t),
\label{eq:S4.Lorentz}
\end{equation}
where \(\eta\) is a damping rate, \(\alpha\) is an effective coupling constant, and \(\kappa(t)\) is a prescribed real \(T\)-periodic modulation. The displacement and magnetic induction are taken as
\begin{equation}
D(z,t) = \varepsilon_v E(z,t) + P(z,t),\qquad
B(z,t) = \mu_v H(z,t),
\label{eq:S4.DB}
\end{equation}
with vacuum parameters \(\varepsilon_v,\mu_v\). Using instead constant background parameters \(\varepsilon_b>0\), \(\mu>0\) in place of \(\varepsilon_v,\mu_v\) is straightforward and does not affect the flux-over-energy structure below. We assume \(\alpha>0\) and \(\kappa(t)\ge \kappa_{\min}>0\) so that the material energy defined below is nonnegative and \(u_{\rm tot}\ge u_{\rm EM}\) pointwise. Maxwell’s equations with a 1D plane-wave ansatz \(E(z,t)=E(t)e^{ikz}\), \(H(z,t)=H(t)e^{ikz}\), \(P(z,t)=P(t)e^{ikz}\) reduce to
\begin{equation}
ik E = -\mu_v\,\partial_t H,
\qquad
ik H = -\partial_t D,
\qquad
\ddot P + \eta\,\dot P + \kappa(t)\,P = \alpha\,E.
\label{eq:S4.Maxwell_Lorentz}
\end{equation}

\noindent
Let \(E_{\rm phys},H_{\rm phys},P_{\rm phys}\) denote the real physical fields.
The instantaneous electromagnetic energy density and flux are
\begin{equation}
u_{\rm EM} = \frac12\bigl(\varepsilon_v E_{\rm phys}^2+\mu_v H_{\rm phys}^2\bigr),
\qquad
S_{z,{\rm phys}} = E_{\rm phys} H_{\rm phys},
\end{equation}
and the medium carries its own internal (oscillator) energy~\cite{SM-LandauLifshitzECM},
\begin{equation}
u_{\rm mat}
= \frac{1}{2\alpha}\Bigl((\dot P_{\rm phys})^2 + \kappa(t)P_{\rm phys}^2\Bigr),
\label{eq:S4.u_mat}
\end{equation}
so that the \emph{total} energy density is
\begin{equation}
u_{\rm tot} = u_{\rm EM} + u_{\rm mat}.
\label{eq:S4.u_tot_def}
\end{equation}
A straightforward calculation, obtained by multiplying the Lorentz equation by \(\dot P_{\rm phys}/\alpha\) and using the real-field Poynting theorem, gives the energy-balance relation
\begin{equation}
\partial_t u_{\rm tot} + \partial_z S_{z,{\rm phys}}
= -\,q_{\rm diss}(z,t) + q_{\rm pump}(z,t),
\label{eq:S4.energy_balance}
\end{equation}
with
\begin{equation}
q_{\rm diss}(z,t) = \eta\,\frac{(\dot P_{\rm phys})^2}{\alpha}\ge 0,
\qquad
q_{\rm pump}(z,t) = \frac{1}{2\alpha}\dot{\kappa}(t)\,P_{\rm phys}^2.
\end{equation}
In the complex notation used below, the same identities hold upon replacing squares by moduli and products by real parts; in particular we use
\(u_{\rm EM}=\tfrac14(\varepsilon_v|E|^2+\mu_v|H|^2)\),
\(S_z=\frac12\Re(EH^*)\),
\(u_{\rm mat}=\tfrac{1}{4\alpha}(|\dot P|^2+\kappa(t)|P|^2)\),
and \(u_{\rm tot}=u_{\rm EM}+u_{\rm mat}\).

\subsection{Lossless dispersive case (\texorpdfstring{$\eta=0$}{eta=0})}
\label{subsec:lorentz_lossless}

\noindent
When \(\eta=0\) the medium is lossless, yet energy can still be exchanged with the external modulation through the time-periodic stiffness \(\kappa(t)\).
Introducing the auxiliary variable \(Q\equiv \dot P\), the Maxwell--Lorentz equations for a 1D plane wave \(\propto e^{ikz}\) can be written in first-order form
\begin{equation}
i\frac{d}{dt}\,|\Psi(t)\rangle = \mathsf H_{\rm ML}(t;k)\,|\Psi(t)\rangle,
\qquad
|\Psi\rangle=(E,H,P,Q)^{\mathsf T},
\label{eq:S4.Schro}
\end{equation}
with
\begin{equation}
\mathsf H_{\rm ML}(t;k)=
\begin{pmatrix}
0 & \dfrac{k}{\varepsilon_v} & 0 & -\dfrac{i}{\varepsilon_v}\\[4pt]
\dfrac{k}{\mu_v} & 0 & 0 & 0\\[4pt]
0 & 0 & 0 & i\\[4pt]
i\alpha & 0 & -i\kappa(t) & 0
\end{pmatrix},
\qquad
\kappa(t+T)=\kappa(t).
\label{eq:S4.Ht}
\end{equation}

\noindent
Passing to the Floquet representation \( |\Psi(t)\rangle=e^{-i\omega t}|\Phi(t)\rangle \) with \( |\Phi(t+T)\rangle=|\Phi(t)\rangle \), we expand each component in temporal harmonics and collect the Fourier coefficients into an extended Floquet vector
\begin{equation}
|R\rangle=\big(\{E_m\},\{H_m\},\{P_m\},\{Q_m\}\big)^{\mathsf T}.
\end{equation}
Defining the Floquet matrices
\begin{equation}
(\Lambda)_{mn}=(\omega+m\Omega)\delta_{mn},\qquad
K=k\,\mathbb I,\qquad
(\mathcal K)_{mn}=\kappa_{m-n},
\quad
\kappa(t)=\sum_\ell \kappa_\ell e^{-i\ell\Omega t},
\label{eq:S4.Sambe_ops}
\end{equation}
the extended constraint matrix \(M_{\rm ext}(\omega,k)\) reads
\begin{equation}
M_{\rm ext}(\omega,k)=
\begin{pmatrix}
K & -\mu_v\Lambda & 0 & 0\\
-\varepsilon_v\Lambda & K & 0 & -i\mathbb I\\
0 & 0 & -i\Lambda & -\mathbb I\\
-\alpha\mathbb I & 0 & \mathcal K & -i\Lambda
\end{pmatrix},
\qquad
M_{\rm ext}(\omega,k)|R\rangle=0.
\label{eq:S4.Mext_block_Q}
\end{equation}

\noindent
We introduce an extended matrix \(\mathbb J_{\rm ext}\) chosen so that the \(\mathbb J_{\rm ext}\)-weighted forms
\(\langle R|\mathbb J_{\rm ext}\partial_k M_{\rm ext}|R\rangle\) and
\(\langle R|\mathbb J_{\rm ext}\partial_\omega M_{\rm ext}|R\rangle\)
reproduce, respectively, the cycle-averaged Poynting flux and the cycle-averaged \emph{total} (field-plus-oscillator) energy density, generalizing the nondispersive construction:
\begin{equation}
\mathbb J_{\rm ext}=
\begin{pmatrix}
0 & \mathbb I & 0 & 0\\
\mathbb I & 0 & 0 & 0\\
0 & 0 & -\dfrac{i}{\alpha}\mathcal K & 0\\
0 & 0 & 0 & -\dfrac{i}{\alpha}\mathbb I
\end{pmatrix}.
\label{eq:S4.Jext}
\end{equation}

\noindent
Since \(K=k\mathbb I\), one has \(\partial_k M_{\rm ext}=\mathrm{diag}(\mathbb I,\mathbb I,0,0)\), hence
\begin{align}
\langle R|\mathbb J_{\rm ext}\partial_k M_{\rm ext}|R\rangle
&=\bm{E}^\dagger\bm{H}+\bm{H}^\dagger\bm{E}
=2\,\Re(\bm{E}^\dagger\bm{H})
=4\,\langle S_z\rangle_T,
\end{align}
where \(\bm{E}=(\{E_m\})^{\mathsf T}\) and \(\bm{H}=(\{H_m\})^{\mathsf T}\). Next, using \(\partial_\omega\Lambda=\mathbb I\) we obtain
\begin{equation}
\partial_\omega M_{\rm ext}=
\begin{pmatrix}
0 & -\mu_v\mathbb I & 0 & 0\\
-\varepsilon_v\mathbb I & 0 & 0 & 0\\
0 & 0 & -i\mathbb I & 0\\
0 & 0 & 0 & -i\mathbb I
\end{pmatrix},
\label{eq:S4.domegaMext}
\end{equation}
and therefore
\begin{align}
\langle R|\mathbb J_{\rm ext}\partial_\omega M_{\rm ext}|R\rangle
&=
-\varepsilon_v\,\bm{E}^\dagger\bm{E}
-\mu_v\,\bm{H}^\dagger\bm{H}
-\frac{1}{\alpha}\bm{Q}^\dagger\bm{Q}
-\frac{1}{\alpha}\bm{P}^\dagger\mathcal K\,\bm{P} \notag\\
&= -4\,\langle u_{\rm tot}\rangle_T,
\label{eq:S4.bilinears_ext}
\end{align}
with \(\bm{P}=(\{P_m\})^{\mathsf T}\), \(\bm{Q}=(\{Q_m\})^{\mathsf T}\).
Thus, in the lossless dispersive case the extended flux-over-energy velocity
\begin{equation}
v_E^{\rm (tot)}(k)
\equiv
-\frac{\langle R|\mathbb J_{\rm ext}\partial_k M_{\rm ext}|R\rangle}
      {\langle R|\mathbb J_{\rm ext}\partial_\omega M_{\rm ext}|R\rangle}
= \frac{\langle S_z\rangle_T}{\langle u_{\rm tot}\rangle_T}
\label{eq:S4.vE_tot}
\end{equation}
coincides exactly with the ratio of the cycle-averaged flux to the cycle-averaged \emph{total} (field-plus-matter) energy density.
In regimes where the oscillator contribution to \(\langle u_{\rm tot}\rangle_T\) is negligible (e.g., weak coupling or sufficiently far from resonance), Eq.~\eqref{eq:S4.vE_tot} reduces to the main-text flux-over-energy expression.

\subsection{Inclusion of loss and generalized velocity}
\label{subsec:lorentz_loss}

\noindent
When damping is present (\(\eta>0\)), the Maxwell--Lorentz model obeys the same local continuity law for the \emph{total} energy (field plus oscillator),
\begin{equation}
\partial_t u_{\rm tot}+\partial_z S_{z,{\rm phys}}=-q_{\rm diss}+q_{\rm pump},
\label{eq:S4.energy_balance_lossy}
\end{equation}
with \(q_{\rm diss}\ge 0\) and \(q_{\rm pump}\) accounting for modulation-induced work.
For a \emph{single} Floquet eigenmode in a spatially uniform medium, the quadratic densities \(u_{\rm tot}(t)\) and \(S_z(t)\) inherit the temporal periodicity of the drive up to an overall exponential envelope set by the Floquet exponent.
In general the quasifrequency is complex, \(\omega=\omega_r+i\gamma\), so that \(u_{\rm tot}\) and \(S_z\) scale as \(e^{2\gamma t}\) times a \(T\)-periodic function.
It is therefore natural to factor out the envelope and work with weighted quantities \(\tilde u_{\rm tot}=e^{-2\gamma t}u_{\rm tot}\) and \(\tilde S_z=e^{-2\gamma t}S_z\), which are strictly \(T\)-periodic for a single mode.
We define weighted period averages
\begin{equation}
\langle\!\langle f\rangle\!\rangle \equiv \frac{1}{T}\int_0^T e^{-2\gamma t} f(t)\,dt,
\end{equation}
which are independent of the cycle origin for a single eigenmode, and introduce the generalized transport velocity
\begin{equation}
v_{\rm gen}^{\rm (tot)}(k)\equiv
\frac{\langle\!\langle S_z\rangle\!\rangle}{\langle\!\langle u_{\rm tot}\rangle\!\rangle}.
\label{eq:S4.vgen_tot}
\end{equation}

\noindent
Finally, the transport velocity remains bounded.
Pointwise one has \(|S_z|\le u_{\rm EM}/\sqrt{\varepsilon_v\mu_v}\), and under the positivity assumptions \(\alpha>0\), \(\kappa(t)\ge \kappa_{\min}>0\) we have \(u_{\rm tot}\ge u_{\rm EM}\).
It follows that
\begin{equation}
\left|\frac{S_z}{u_{\rm tot}}\right|\le \frac{1}{\sqrt{\varepsilon_v\mu_v}} \equiv c.
\label{eq:S4.v_bound_tot}
\end{equation}
Thus, even with dispersion, weak loss, and modulation pumping, the physically relevant transport speed is naturally characterized by a flux-over-total-energy ratio (or its weighted generalization in the presence of exponential growth/decay), and it remains finite and bounded by the vacuum phase-velocity ceiling.

\bigskip

\makeatletter
\renewcommand\bibsection{%
  \begin{center}
    \textbf{Supplementary References}
  \end{center}
  \vspace{0.5\baselineskip}
}
\makeatother
\begingroup
\makeatletter
\let\SM@origlabel\label
\renewcommand{\label}[1]{%
  \def\SM@tempa{#1}%
  \def\SM@tempb{LastBibItem}%
  \ifx\SM@tempa\SM@tempb
    \SM@origlabel{LastSuppBibItem}%
  \else
    \SM@origlabel{#1}%
  \fi
}
\makeatother

\endgroup


\begin{thebibliography}{47}%
\makeatletter
\providecommand \@ifxundefined [1]{%
 \@ifx{#1\undefined}
}%
\providecommand \@ifnum [1]{%
 \ifnum #1\expandafter \@firstoftwo
 \else \expandafter \@secondoftwo
 \fi
}%
\providecommand \@ifx [1]{%
 \ifx #1\expandafter \@firstoftwo
 \else \expandafter \@secondoftwo
 \fi
}%
\providecommand \natexlab [1]{#1}%
\providecommand \enquote  [1]{``#1''}%
\providecommand \bibnamefont  [1]{#1}%
\providecommand \bibfnamefont [1]{#1}%
\providecommand \citenamefont [1]{#1}%
\providecommand \href@noop [0]{\@secondoftwo}%
\providecommand \href [0]{\begingroup \@sanitize@url \@href}%
\providecommand \@href[1]{\@@startlink{#1}\@@href}%
\providecommand \@@href[1]{\endgroup#1\@@endlink}%
\providecommand \@sanitize@url [0]{\catcode `\\12\catcode `\$12\catcode
  `\&12\catcode `\#12\catcode `\^12\catcode `\_12\catcode `\%12\relax}%
\providecommand \@@startlink[1]{}%
\providecommand \@@endlink[0]{}%
\providecommand \url  [0]{\begingroup\@sanitize@url \@url }%
\providecommand \@url [1]{\endgroup\@href {#1}{\urlprefix }}%
\providecommand \urlprefix  [0]{URL }%
\providecommand \Eprint [0]{\href }%
\providecommand \doibase [0]{https://doi.org/}%
\providecommand \selectlanguage [0]{\@gobble}%
\providecommand \bibinfo  [0]{\@secondoftwo}%
\providecommand \bibfield  [0]{\@secondoftwo}%
\providecommand \translation [1]{[#1]}%
\providecommand \BibitemOpen [0]{}%
\providecommand \bibitemStop [0]{}%
\providecommand \bibitemNoStop [0]{.\EOS\space}%
\providecommand \EOS [0]{\spacefactor3000\relax}%
\providecommand \BibitemShut  [1]{\csname bibitem#1\endcsname}%
\let\auto@bib@innerbib\@empty
\bibitem [{\citenamefont {Sommerfeld}(1914)}]{sommerfeld1914fortpflanzung}%
  \BibitemOpen
  \bibfield  {author} {\bibinfo {author} {\bibfnamefont {A.}~\bibnamefont
  {Sommerfeld}},\ }\bibfield  {title} {\bibinfo {title} {{\"U}ber die
  fortpflanzung des lichtes in dispergierenden medien},\ }\href
  {https://doi.org/10.1002/andp.19143491002} {\bibfield  {journal} {\bibinfo
  {journal} {Annalen der Physik}\ }\textbf {\bibinfo {volume} {349}},\ \bibinfo
  {pages} {177} (\bibinfo {year} {1914})}\BibitemShut {NoStop}%
\bibitem [{\citenamefont {Brillouin}(2013)}]{brillouin2013wave}%
  \BibitemOpen
  \bibfield  {author} {\bibinfo {author} {\bibfnamefont {L.}~\bibnamefont
  {Brillouin}},\ }\href
  {https://shop.elsevier.com/books/wave-propagation-and-group-velocity/brillouin/978-1-4832-3068-9}
  {\emph {\bibinfo {title} {Wave Propagation and Group Velocity}}},\ \bibinfo
  {edition} {revised}\ ed.,\ edited by\ \bibinfo {editor} {\bibfnamefont
  {H.~S.~W.}\ \bibnamefont {Massey}},\ \bibinfo {series} {Pure and Applied
  Physics}, Vol.~\bibinfo {volume} {8}\ (\bibinfo  {publisher} {Academic
  Press},\ \bibinfo {address} {New York, NY},\ \bibinfo {year} {2013})\
  \bibinfo {note} {reissue of the classic 1960 volume; print ISBN:
  9781483230689}\BibitemShut {NoStop}%
\bibitem [{\citenamefont {Wang}\ \emph {et~al.}(2000)\citenamefont {Wang},
  \citenamefont {Kuzmich},\ and\ \citenamefont {Dogariu}}]{wang2000gain}%
  \BibitemOpen
  \bibfield  {author} {\bibinfo {author} {\bibfnamefont {L.~J.}\ \bibnamefont
  {Wang}}, \bibinfo {author} {\bibfnamefont {A.}~\bibnamefont {Kuzmich}},\ and\
  \bibinfo {author} {\bibfnamefont {A.}~\bibnamefont {Dogariu}},\ }\bibfield
  {title} {\bibinfo {title} {Gain-assisted superluminal light propagation},\
  }\href {https://doi.org/10.1038/35018520} {\bibfield  {journal} {\bibinfo
  {journal} {Nature}\ }\textbf {\bibinfo {volume} {406}},\ \bibinfo {pages}
  {277} (\bibinfo {year} {2000})}\BibitemShut {NoStop}%
\bibitem [{\citenamefont {Oughstun}\ and\ \citenamefont
  {Sherman}(1994)}]{oughstun2012electromagnetic}%
  \BibitemOpen
  \bibfield  {author} {\bibinfo {author} {\bibfnamefont {K.~E.}\ \bibnamefont
  {Oughstun}}\ and\ \bibinfo {author} {\bibfnamefont {G.~C.}\ \bibnamefont
  {Sherman}},\ }\href {https://doi.org/10.1007/978-3-642-61227-5} {\emph
  {\bibinfo {title} {Electromagnetic Pulse Propagation in Causal
  Dielectrics}}},\ \bibinfo {edition} {1st}\ ed.,\ \bibinfo {series} {Springer
  Series on Wave Phenomena}, Vol.~\bibinfo {volume} {16}\ (\bibinfo
  {publisher} {Springer},\ \bibinfo {address} {Berlin, Heidelberg},\ \bibinfo
  {year} {1994})\ \bibinfo {note} {eBook ISBN: 9783642612275}\BibitemShut
  {NoStop}%
\bibitem [{\citenamefont {Peatross}\ \emph {et~al.}(2000)\citenamefont
  {Peatross}, \citenamefont {Glasgow},\ and\ \citenamefont
  {Ware}}]{peatross2000average}%
  \BibitemOpen
  \bibfield  {author} {\bibinfo {author} {\bibfnamefont {J.}~\bibnamefont
  {Peatross}}, \bibinfo {author} {\bibfnamefont {S.~A.}\ \bibnamefont
  {Glasgow}},\ and\ \bibinfo {author} {\bibfnamefont {M.}~\bibnamefont
  {Ware}},\ }\bibfield  {title} {\bibinfo {title} {Average energy flow of
  optical pulses in dispersive media},\ }\href
  {https://doi.org/10.1103/PhysRevLett.84.2370} {\bibfield  {journal} {\bibinfo
   {journal} {Physical Review Letters}\ }\textbf {\bibinfo {volume} {84}},\
  \bibinfo {pages} {2370} (\bibinfo {year} {2000})}\BibitemShut {NoStop}%
\bibitem [{\citenamefont {Milonni}(2004)}]{milonni2004fast}%
  \BibitemOpen
  \bibfield  {author} {\bibinfo {author} {\bibfnamefont {P.~W.}\ \bibnamefont
  {Milonni}},\ }\href
  {https://www.routledge.com/Fast-Light-Slow-Light-and-Left-Handed-Light/Milonni/p/book/9780367578206}
  {\emph {\bibinfo {title} {Fast Light, Slow Light and Left-Handed Light}}},\
  \bibinfo {edition} {1st}\ ed.,\ Series in Optics and Optoelectronics\
  (\bibinfo  {publisher} {CRC Press},\ \bibinfo {address} {Boca Raton, FL},\
  \bibinfo {year} {2004})\ \bibinfo {note} {cRC Press is an imprint of Taylor
  \& Francis}\BibitemShut {NoStop}%
\bibitem [{\citenamefont {Boyd}\ and\ \citenamefont
  {Gauthier}(2009)}]{boyd2009controlling}%
  \BibitemOpen
  \bibfield  {author} {\bibinfo {author} {\bibfnamefont {R.~W.}\ \bibnamefont
  {Boyd}}\ and\ \bibinfo {author} {\bibfnamefont {D.~J.}\ \bibnamefont
  {Gauthier}},\ }\bibfield  {title} {\bibinfo {title} {Controlling the velocity
  of light pulses},\ }\href {https://doi.org/10.1126/science.1170885}
  {\bibfield  {journal} {\bibinfo  {journal} {Science}\ }\textbf {\bibinfo
  {volume} {326}},\ \bibinfo {pages} {1074} (\bibinfo {year}
  {2009})}\BibitemShut {NoStop}%
\bibitem [{\citenamefont {Huang}\ and\ \citenamefont
  {Zhang}(2001)}]{huang2001poynting}%
  \BibitemOpen
  \bibfield  {author} {\bibinfo {author} {\bibfnamefont {C.-G.}\ \bibnamefont
  {Huang}}\ and\ \bibinfo {author} {\bibfnamefont {Y.-Z.}\ \bibnamefont
  {Zhang}},\ }\bibfield  {title} {\bibinfo {title} {Poynting vector, energy
  density, and energy velocity in an anomalous dispersion medium},\ }\href
  {https://doi.org/10.1103/PhysRevA.65.015802} {\bibfield  {journal} {\bibinfo
  {journal} {Physical Review A}\ }\textbf {\bibinfo {volume} {65}},\ \bibinfo
  {pages} {015802} (\bibinfo {year} {2001})}\BibitemShut {NoStop}%
\bibitem [{\citenamefont {Kuzmich}\ \emph {et~al.}(2001)\citenamefont
  {Kuzmich}, \citenamefont {Dogariu}, \citenamefont {Wang}, \citenamefont
  {Milonni},\ and\ \citenamefont {Chiao}}]{kuzmich2001signal}%
  \BibitemOpen
  \bibfield  {author} {\bibinfo {author} {\bibfnamefont {A.}~\bibnamefont
  {Kuzmich}}, \bibinfo {author} {\bibfnamefont {A.}~\bibnamefont {Dogariu}},
  \bibinfo {author} {\bibfnamefont {L.~J.}\ \bibnamefont {Wang}}, \bibinfo
  {author} {\bibfnamefont {P.~W.}\ \bibnamefont {Milonni}},\ and\ \bibinfo
  {author} {\bibfnamefont {R.~Y.}\ \bibnamefont {Chiao}},\ }\bibfield  {title}
  {\bibinfo {title} {Signal velocity, causality, and quantum noise in
  superluminal light pulse propagation},\ }\href
  {https://doi.org/10.1103/PhysRevLett.86.3925} {\bibfield  {journal} {\bibinfo
   {journal} {Physical Review Letters}\ }\textbf {\bibinfo {volume} {86}},\
  \bibinfo {pages} {3925} (\bibinfo {year} {2001})}\BibitemShut {NoStop}%
\bibitem [{\citenamefont {Harris}\ \emph {et~al.}(1990)\citenamefont {Harris},
  \citenamefont {Field},\ and\ \citenamefont
  {Imamo{\u{g}}lu}}]{PhysRevLett.64.1107}%
  \BibitemOpen
  \bibfield  {author} {\bibinfo {author} {\bibfnamefont {S.~E.}\ \bibnamefont
  {Harris}}, \bibinfo {author} {\bibfnamefont {J.~E.}\ \bibnamefont {Field}},\
  and\ \bibinfo {author} {\bibfnamefont {A.}~\bibnamefont {Imamo{\u{g}}lu}},\
  }\bibfield  {title} {\bibinfo {title} {Nonlinear optical processes using
  electromagnetically induced transparency},\ }\href
  {https://doi.org/10.1103/PhysRevLett.64.1107} {\bibfield  {journal} {\bibinfo
   {journal} {Physical Review Letters}\ }\textbf {\bibinfo {volume} {64}},\
  \bibinfo {pages} {1107} (\bibinfo {year} {1990})}\BibitemShut {NoStop}%
\bibitem [{\citenamefont {Hau}\ \emph {et~al.}(1999)\citenamefont {Hau},
  \citenamefont {Harris}, \citenamefont {Dutton},\ and\ \citenamefont
  {Behroozi}}]{hau1999light}%
  \BibitemOpen
  \bibfield  {author} {\bibinfo {author} {\bibfnamefont {L.~V.}\ \bibnamefont
  {Hau}}, \bibinfo {author} {\bibfnamefont {S.~E.}\ \bibnamefont {Harris}},
  \bibinfo {author} {\bibfnamefont {Z.}~\bibnamefont {Dutton}},\ and\ \bibinfo
  {author} {\bibfnamefont {C.~H.}\ \bibnamefont {Behroozi}},\ }\bibfield
  {title} {\bibinfo {title} {Light speed reduction to 17 metres per second in
  an ultracold atomic gas},\ }\href {https://doi.org/10.1038/17561} {\bibfield
  {journal} {\bibinfo  {journal} {Nature}\ }\textbf {\bibinfo {volume} {397}},\
  \bibinfo {pages} {594} (\bibinfo {year} {1999})}\BibitemShut {NoStop}%
\bibitem [{\citenamefont {Baba}(2008)}]{baba2008slow}%
  \BibitemOpen
  \bibfield  {author} {\bibinfo {author} {\bibfnamefont {T.}~\bibnamefont
  {Baba}},\ }\bibfield  {title} {\bibinfo {title} {Slow light in photonic
  crystals},\ }\href {https://doi.org/10.1038/nphoton.2008.146} {\bibfield
  {journal} {\bibinfo  {journal} {Nature Photonics}\ }\textbf {\bibinfo
  {volume} {2}},\ \bibinfo {pages} {465} (\bibinfo {year} {2008})}\BibitemShut
  {NoStop}%
\bibitem [{\citenamefont {D'Aguanno}\ \emph {et~al.}(2001)\citenamefont
  {D'Aguanno}, \citenamefont {Centini}, \citenamefont {Scalora}, \citenamefont
  {Sibilia}, \citenamefont {Bloemer}, \citenamefont {Bowden}, \citenamefont
  {Haus},\ and\ \citenamefont {Bertolotti}}]{d2001group}%
  \BibitemOpen
  \bibfield  {author} {\bibinfo {author} {\bibfnamefont {G.}~\bibnamefont
  {D'Aguanno}}, \bibinfo {author} {\bibfnamefont {M.}~\bibnamefont {Centini}},
  \bibinfo {author} {\bibfnamefont {M.}~\bibnamefont {Scalora}}, \bibinfo
  {author} {\bibfnamefont {C.}~\bibnamefont {Sibilia}}, \bibinfo {author}
  {\bibfnamefont {M.~J.}\ \bibnamefont {Bloemer}}, \bibinfo {author}
  {\bibfnamefont {C.~M.}\ \bibnamefont {Bowden}}, \bibinfo {author}
  {\bibfnamefont {J.~W.}\ \bibnamefont {Haus}},\ and\ \bibinfo {author}
  {\bibfnamefont {M.}~\bibnamefont {Bertolotti}},\ }\bibfield  {title}
  {\bibinfo {title} {Group velocity, energy velocity, and superluminal
  propagation in finite photonic band-gap structures},\ }\href
  {https://doi.org/10.1103/PhysRevE.63.036610} {\bibfield  {journal} {\bibinfo
  {journal} {Physical Review E}\ }\textbf {\bibinfo {volume} {63}},\ \bibinfo
  {pages} {036610} (\bibinfo {year} {2001})}\BibitemShut {NoStop}%
\bibitem [{\citenamefont {Garrett}\ and\ \citenamefont
  {McCumber}(1970)}]{garrett1970propagation}%
  \BibitemOpen
  \bibfield  {author} {\bibinfo {author} {\bibfnamefont {C.~G.~B.}\
  \bibnamefont {Garrett}}\ and\ \bibinfo {author} {\bibfnamefont {D.~E.}\
  \bibnamefont {McCumber}},\ }\bibfield  {title} {\bibinfo {title} {Propagation
  of a {G}aussian light pulse through an anomalous dispersion medium},\ }\href
  {https://doi.org/10.1103/PhysRevA.1.305} {\bibfield  {journal} {\bibinfo
  {journal} {Physical Review A}\ }\textbf {\bibinfo {volume} {1}},\ \bibinfo
  {pages} {305} (\bibinfo {year} {1970})}\BibitemShut {NoStop}%
\bibitem [{\citenamefont {Chu}\ and\ \citenamefont
  {Wong}(1982)}]{chu1982linear}%
  \BibitemOpen
  \bibfield  {author} {\bibinfo {author} {\bibfnamefont {S.}~\bibnamefont
  {Chu}}\ and\ \bibinfo {author} {\bibfnamefont {S.}~\bibnamefont {Wong}},\
  }\bibfield  {title} {\bibinfo {title} {Linear pulse propagation in an
  absorbing medium},\ }\href {https://doi.org/10.1103/PhysRevLett.48.738}
  {\bibfield  {journal} {\bibinfo  {journal} {Physical Review Letters}\
  }\textbf {\bibinfo {volume} {48}},\ \bibinfo {pages} {738} (\bibinfo {year}
  {1982})}\BibitemShut {NoStop}%
\bibitem [{\citenamefont {Stenner}\ \emph {et~al.}(2003)\citenamefont
  {Stenner}, \citenamefont {Gauthier},\ and\ \citenamefont
  {Neifeld}}]{stenner2003speed}%
  \BibitemOpen
  \bibfield  {author} {\bibinfo {author} {\bibfnamefont {M.~D.}\ \bibnamefont
  {Stenner}}, \bibinfo {author} {\bibfnamefont {D.~J.}\ \bibnamefont
  {Gauthier}},\ and\ \bibinfo {author} {\bibfnamefont {M.~A.}\ \bibnamefont
  {Neifeld}},\ }\bibfield  {title} {\bibinfo {title} {The speed of information
  in a `fast-light' optical medium},\ }\href
  {https://doi.org/10.1038/nature02016} {\bibfield  {journal} {\bibinfo
  {journal} {Nature}\ }\textbf {\bibinfo {volume} {425}},\ \bibinfo {pages}
  {695} (\bibinfo {year} {2003})}\BibitemShut {NoStop}%
\bibitem [{\citenamefont {Caloz}\ and\ \citenamefont
  {Deck-L{\'e}ger}(2020)}]{caloz2019spacetime}%
  \BibitemOpen
  \bibfield  {author} {\bibinfo {author} {\bibfnamefont {C.}~\bibnamefont
  {Caloz}}\ and\ \bibinfo {author} {\bibfnamefont {Z.-L.}\ \bibnamefont
  {Deck-L{\'e}ger}},\ }\bibfield  {title} {\bibinfo {title} {Spacetime
  metamaterials---part {I}: general concepts},\ }\href
  {https://doi.org/10.1109/TAP.2019.2944225} {\bibfield  {journal} {\bibinfo
  {journal} {IEEE Transactions on Antennas and Propagation}\ }\textbf {\bibinfo
  {volume} {68}},\ \bibinfo {pages} {1569} (\bibinfo {year}
  {2020})}\BibitemShut {NoStop}%
\bibitem [{\citenamefont {Park}\ and\ \citenamefont
  {Min}(2021)}]{park2021spatiotemporal}%
  \BibitemOpen
  \bibfield  {author} {\bibinfo {author} {\bibfnamefont {J.}~\bibnamefont
  {Park}}\ and\ \bibinfo {author} {\bibfnamefont {B.}~\bibnamefont {Min}},\
  }\bibfield  {title} {\bibinfo {title} {Spatiotemporal plane wave expansion
  method for arbitrary space--time periodic photonic media},\ }\href
  {https://doi.org/10.1364/OL.411622} {\bibfield  {journal} {\bibinfo
  {journal} {Optics Letters}\ }\textbf {\bibinfo {volume} {46}},\ \bibinfo
  {pages} {484} (\bibinfo {year} {2021})}\BibitemShut {NoStop}%
\bibitem [{\citenamefont {Galiffi}\ \emph {et~al.}(2022)\citenamefont
  {Galiffi}, \citenamefont {Tirole}, \citenamefont {Yin}, \citenamefont {Li},
  \citenamefont {Vezzoli}, \citenamefont {Huidobro}, \citenamefont
  {Silveirinha}, \citenamefont {Sapienza}, \citenamefont {Al{\`u}},\ and\
  \citenamefont {Pendry}}]{galiffi2022photonics}%
  \BibitemOpen
  \bibfield  {author} {\bibinfo {author} {\bibfnamefont {E.}~\bibnamefont
  {Galiffi}}, \bibinfo {author} {\bibfnamefont {R.}~\bibnamefont {Tirole}},
  \bibinfo {author} {\bibfnamefont {S.}~\bibnamefont {Yin}}, \bibinfo {author}
  {\bibfnamefont {H.}~\bibnamefont {Li}}, \bibinfo {author} {\bibfnamefont
  {S.}~\bibnamefont {Vezzoli}}, \bibinfo {author} {\bibfnamefont {P.~A.}\
  \bibnamefont {Huidobro}}, \bibinfo {author} {\bibfnamefont {M.~G.}\
  \bibnamefont {Silveirinha}}, \bibinfo {author} {\bibfnamefont
  {R.}~\bibnamefont {Sapienza}}, \bibinfo {author} {\bibfnamefont
  {A.}~\bibnamefont {Al{\`u}}},\ and\ \bibinfo {author} {\bibfnamefont {J.~B.}\
  \bibnamefont {Pendry}},\ }\bibfield  {title} {\bibinfo {title} {Photonics of
  time-varying media},\ }\href {https://doi.org/10.1117/1.AP.4.1.014002}
  {\bibfield  {journal} {\bibinfo  {journal} {Advanced Photonics}\ }\textbf
  {\bibinfo {volume} {4}},\ \bibinfo {pages} {014002} (\bibinfo {year}
  {2022})}\BibitemShut {NoStop}%
\bibitem [{\citenamefont {Pacheco-Pe{\~n}a}\ \emph {et~al.}(2022)\citenamefont
  {Pacheco-Pe{\~n}a}, \citenamefont {Sol{\'i}s},\ and\ \citenamefont
  {Engheta}}]{pacheco2022time}%
  \BibitemOpen
  \bibfield  {author} {\bibinfo {author} {\bibfnamefont {V.}~\bibnamefont
  {Pacheco-Pe{\~n}a}}, \bibinfo {author} {\bibfnamefont {D.~M.}\ \bibnamefont
  {Sol{\'i}s}},\ and\ \bibinfo {author} {\bibfnamefont {N.}~\bibnamefont
  {Engheta}},\ }\bibfield  {title} {\bibinfo {title} {Time-varying
  electromagnetic media: opinion},\ }\href {https://doi.org/10.1364/OME.471007}
  {\bibfield  {journal} {\bibinfo  {journal} {Optical Materials Express}\
  }\textbf {\bibinfo {volume} {12}},\ \bibinfo {pages} {3829} (\bibinfo {year}
  {2022})}\BibitemShut {NoStop}%
\bibitem [{\citenamefont {Asgari}\ \emph {et~al.}(2024)\citenamefont {Asgari},
  \citenamefont {Garg}, \citenamefont {Wang}, \citenamefont {Mirmoosa},
  \citenamefont {Rockstuhl},\ and\ \citenamefont {Asadchy}}]{asgari2024theory}%
  \BibitemOpen
  \bibfield  {author} {\bibinfo {author} {\bibfnamefont {M.~M.}\ \bibnamefont
  {Asgari}}, \bibinfo {author} {\bibfnamefont {P.}~\bibnamefont {Garg}},
  \bibinfo {author} {\bibfnamefont {X.}~\bibnamefont {Wang}}, \bibinfo {author}
  {\bibfnamefont {M.~S.}\ \bibnamefont {Mirmoosa}}, \bibinfo {author}
  {\bibfnamefont {C.}~\bibnamefont {Rockstuhl}},\ and\ \bibinfo {author}
  {\bibfnamefont {V.}~\bibnamefont {Asadchy}},\ }\bibfield  {title} {\bibinfo
  {title} {Theory and applications of photonic time crystals: a tutorial},\
  }\href {https://doi.org/10.1364/AOP.525163} {\bibfield  {journal} {\bibinfo
  {journal} {Advances in Optics and Photonics}\ }\textbf {\bibinfo {volume}
  {16}},\ \bibinfo {pages} {958} (\bibinfo {year} {2024})}\BibitemShut
  {NoStop}%
\bibitem [{\citenamefont {Horsley}\ and\ \citenamefont
  {Pendry}(2023)}]{horsley2023quantum}%
  \BibitemOpen
  \bibfield  {author} {\bibinfo {author} {\bibfnamefont {S.~A.~R.}\
  \bibnamefont {Horsley}}\ and\ \bibinfo {author} {\bibfnamefont {J.~B.}\
  \bibnamefont {Pendry}},\ }\bibfield  {title} {\bibinfo {title} {Quantum
  electrodynamics of time-varying gratings},\ }\href
  {https://doi.org/10.1073/pnas.2302652120} {\bibfield  {journal} {\bibinfo
  {journal} {Proceedings of the National Academy of Sciences}\ }\textbf
  {\bibinfo {volume} {120}},\ \bibinfo {pages} {e2302652120} (\bibinfo {year}
  {2023})}\BibitemShut {NoStop}%
\bibitem [{\citenamefont {Pendry}\ and\ \citenamefont
  {Horsley}(2024)}]{pendry2024qed}%
  \BibitemOpen
  \bibfield  {author} {\bibinfo {author} {\bibfnamefont {J.~B.}\ \bibnamefont
  {Pendry}}\ and\ \bibinfo {author} {\bibfnamefont {S.~A.~R.}\ \bibnamefont
  {Horsley}},\ }\bibfield  {title} {\bibinfo {title} {{QED} in space--time
  varying materials},\ }\href {https://doi.org/10.1063/5.0199503} {\bibfield
  {journal} {\bibinfo  {journal} {APL Quantum}\ }\textbf {\bibinfo {volume}
  {1}},\ \bibinfo {pages} {020901} (\bibinfo {year} {2024})}\BibitemShut
  {NoStop}%
\bibitem [{\citenamefont {Park}\ \emph {et~al.}(2025)\citenamefont {Park},
  \citenamefont {Lee}, \citenamefont {Zhang}, \citenamefont {Park},
  \citenamefont {Ryu}, \citenamefont {Cho}, \citenamefont {Lee}, \citenamefont
  {Zhang}, \citenamefont {Park}, \citenamefont {Jeon}, \citenamefont {Shin},
  \citenamefont {Chan},\ and\ \citenamefont {Min}}]{park2025spontaneous}%
  \BibitemOpen
  \bibfield  {author} {\bibinfo {author} {\bibfnamefont {J.}~\bibnamefont
  {Park}}, \bibinfo {author} {\bibfnamefont {K.}~\bibnamefont {Lee}}, \bibinfo
  {author} {\bibfnamefont {R.-Y.}\ \bibnamefont {Zhang}}, \bibinfo {author}
  {\bibfnamefont {H.-C.}\ \bibnamefont {Park}}, \bibinfo {author}
  {\bibfnamefont {J.-W.}\ \bibnamefont {Ryu}}, \bibinfo {author} {\bibfnamefont
  {G.~Y.}\ \bibnamefont {Cho}}, \bibinfo {author} {\bibfnamefont {M.~Y.}\
  \bibnamefont {Lee}}, \bibinfo {author} {\bibfnamefont {Z.}~\bibnamefont
  {Zhang}}, \bibinfo {author} {\bibfnamefont {N.}~\bibnamefont {Park}},
  \bibinfo {author} {\bibfnamefont {W.}~\bibnamefont {Jeon}}, \bibinfo {author}
  {\bibfnamefont {J.}~\bibnamefont {Shin}}, \bibinfo {author} {\bibfnamefont
  {C.~T.}\ \bibnamefont {Chan}},\ and\ \bibinfo {author} {\bibfnamefont
  {B.}~\bibnamefont {Min}},\ }\bibfield  {title} {\bibinfo {title} {Spontaneous
  emission decay and excitation in photonic time crystals},\ }\href
  {https://doi.org/10.1103/5v2w-yg7v} {\bibfield  {journal} {\bibinfo
  {journal} {Physical Review Letters}\ }\textbf {\bibinfo {volume} {135}},\
  \bibinfo {pages} {133801} (\bibinfo {year} {2025})}\BibitemShut {NoStop}%
\bibitem [{\citenamefont {Bae}\ \emph {et~al.}(2026)\citenamefont {Bae},
  \citenamefont {Lee}, \citenamefont {Min},\ and\ \citenamefont
  {Kim}}]{bae2025quantum}%
  \BibitemOpen
  \bibfield  {author} {\bibinfo {author} {\bibfnamefont {J.}~\bibnamefont
  {Bae}}, \bibinfo {author} {\bibfnamefont {K.}~\bibnamefont {Lee}}, \bibinfo
  {author} {\bibfnamefont {B.}~\bibnamefont {Min}},\ and\ \bibinfo {author}
  {\bibfnamefont {K.~W.}\ \bibnamefont {Kim}},\ }\bibfield  {title} {\bibinfo
  {title} {Quantum electrodynamics of photonic time crystals},\ }\href
  {https://doi.org/10.1038/s41467-025-67572-0} {\bibfield  {journal} {\bibinfo
  {journal} {Nature Communications}\ }\textbf {\bibinfo {volume} {17}},\
  \bibinfo {pages} {858} (\bibinfo {year} {2026})},\ \bibinfo {note} {published
  online 19 Dec 2025}\BibitemShut {NoStop}%
\bibitem [{\citenamefont {Wang}\ \emph {et~al.}(2018)\citenamefont {Wang},
  \citenamefont {Zhang},\ and\ \citenamefont {Chan}}]{PhysRevB.98.085142}%
  \BibitemOpen
  \bibfield  {author} {\bibinfo {author} {\bibfnamefont {N.}~\bibnamefont
  {Wang}}, \bibinfo {author} {\bibfnamefont {Z.-Q.}\ \bibnamefont {Zhang}},\
  and\ \bibinfo {author} {\bibfnamefont {C.~T.}\ \bibnamefont {Chan}},\
  }\bibfield  {title} {\bibinfo {title} {Photonic {F}loquet media with a
  complex time-periodic permittivity},\ }\href
  {https://doi.org/10.1103/PhysRevB.98.085142} {\bibfield  {journal} {\bibinfo
  {journal} {Physical Review B}\ }\textbf {\bibinfo {volume} {98}},\ \bibinfo
  {pages} {085142} (\bibinfo {year} {2018})}\BibitemShut {NoStop}%
\bibitem [{\citenamefont {Sustaeta-Osuna}\ \emph {et~al.}(2025)\citenamefont
  {Sustaeta-Osuna}, \citenamefont {Garc{\'\i}a-Vidal},\ and\ \citenamefont
  {Huidobro}}]{sustaeta2025quantum}%
  \BibitemOpen
  \bibfield  {author} {\bibinfo {author} {\bibfnamefont {J.~E.}\ \bibnamefont
  {Sustaeta-Osuna}}, \bibinfo {author} {\bibfnamefont {F.~J.}\ \bibnamefont
  {Garc{\'\i}a-Vidal}},\ and\ \bibinfo {author} {\bibfnamefont
  {P.}~\bibnamefont {Huidobro}},\ }\bibfield  {title} {\bibinfo {title}
  {Quantum theory of photon pair creation in photonic time crystals},\
  }\href@noop {} {\bibfield  {journal} {\bibinfo  {journal} {ACS photonics}\
  }\textbf {\bibinfo {volume} {12}},\ \bibinfo {pages} {1873} (\bibinfo {year}
  {2025})}\BibitemShut {NoStop}%
\bibitem [{\citenamefont {Allard}\ \emph {et~al.}(2026)\citenamefont {Allard},
  \citenamefont {Sustaeta-Osuna}, \citenamefont {Garc{\'\i}a-Vidal},\ and\
  \citenamefont {Huidobro}}]{allard2026broadband}%
  \BibitemOpen
  \bibfield  {author} {\bibinfo {author} {\bibfnamefont {T.~F.}\ \bibnamefont
  {Allard}}, \bibinfo {author} {\bibfnamefont {J.~E.}\ \bibnamefont
  {Sustaeta-Osuna}}, \bibinfo {author} {\bibfnamefont {F.~J.}\ \bibnamefont
  {Garc{\'\i}a-Vidal}},\ and\ \bibinfo {author} {\bibfnamefont {P.~A.}\
  \bibnamefont {Huidobro}},\ }\bibfield  {title} {\bibinfo {title} {Broadband
  dipole absorption in dispersive photonic time crystals},\ }\href@noop {}
  {\bibfield  {journal} {\bibinfo  {journal} {Physical Review Letters}\
  }\textbf {\bibinfo {volume} {136}},\ \bibinfo {pages} {106903} (\bibinfo
  {year} {2026})}\BibitemShut {NoStop}%
\bibitem [{\citenamefont {Shirley}(1965)}]{shirley1965solution}%
  \BibitemOpen
  \bibfield  {author} {\bibinfo {author} {\bibfnamefont {J.~H.}\ \bibnamefont
  {Shirley}},\ }\bibfield  {title} {\bibinfo {title} {Solution of the
  {S}chr{\"o}dinger equation with a {H}amiltonian periodic in time},\ }\href
  {https://doi.org/10.1103/PhysRev.138.B979} {\bibfield  {journal} {\bibinfo
  {journal} {Physical Review}\ }\textbf {\bibinfo {volume} {138}},\ \bibinfo
  {pages} {B979} (\bibinfo {year} {1965})}\BibitemShut {NoStop}%
\bibitem [{\citenamefont {Sambe}(1973)}]{sambe1973steady}%
  \BibitemOpen
  \bibfield  {author} {\bibinfo {author} {\bibfnamefont {H.}~\bibnamefont
  {Sambe}},\ }\bibfield  {title} {\bibinfo {title} {Steady states and
  quasienergies of a quantum-mechanical system in an oscillating field},\
  }\href {https://doi.org/10.1103/PhysRevA.7.2203} {\bibfield  {journal}
  {\bibinfo  {journal} {Physical Review A}\ }\textbf {\bibinfo {volume} {7}},\
  \bibinfo {pages} {2203} (\bibinfo {year} {1973})}\BibitemShut {NoStop}%
\bibitem [{\citenamefont {Tien}(1958)}]{tien1958parametric}%
  \BibitemOpen
  \bibfield  {author} {\bibinfo {author} {\bibfnamefont {P.~K.}\ \bibnamefont
  {Tien}},\ }\bibfield  {title} {\bibinfo {title} {Parametric amplification and
  frequency mixing in propagating circuits},\ }\href
  {https://doi.org/10.1063/1.1723440} {\bibfield  {journal} {\bibinfo
  {journal} {Journal of Applied Physics}\ }\textbf {\bibinfo {volume} {29}},\
  \bibinfo {pages} {1347} (\bibinfo {year} {1958})}\BibitemShut {NoStop}%
\bibitem [{\citenamefont {Lee}\ \emph {et~al.}(2021)\citenamefont {Lee},
  \citenamefont {Park}, \citenamefont {Cho}, \citenamefont {Wang},
  \citenamefont {Kim}, \citenamefont {Daraio},\ and\ \citenamefont
  {Min}}]{lee2021parametric}%
  \BibitemOpen
  \bibfield  {author} {\bibinfo {author} {\bibfnamefont {S.}~\bibnamefont
  {Lee}}, \bibinfo {author} {\bibfnamefont {J.}~\bibnamefont {Park}}, \bibinfo
  {author} {\bibfnamefont {H.}~\bibnamefont {Cho}}, \bibinfo {author}
  {\bibfnamefont {Y.}~\bibnamefont {Wang}}, \bibinfo {author} {\bibfnamefont
  {B.}~\bibnamefont {Kim}}, \bibinfo {author} {\bibfnamefont {C.}~\bibnamefont
  {Daraio}},\ and\ \bibinfo {author} {\bibfnamefont {B.}~\bibnamefont {Min}},\
  }\bibfield  {title} {\bibinfo {title} {Parametric oscillation of
  electromagnetic waves in momentum band gaps of a spatiotemporal crystal},\
  }\href {https://doi.org/10.1364/PRJ.406215} {\bibfield  {journal} {\bibinfo
  {journal} {Photonics Research}\ }\textbf {\bibinfo {volume} {9}},\ \bibinfo
  {pages} {142} (\bibinfo {year} {2021})}\BibitemShut {NoStop}%
\bibitem [{\citenamefont {Park}\ \emph {et~al.}(2022)\citenamefont {Park},
  \citenamefont {Cho}, \citenamefont {Lee}, \citenamefont {Lee}, \citenamefont
  {Lee}, \citenamefont {Park}, \citenamefont {Ryu}, \citenamefont {Park},
  \citenamefont {Jeon},\ and\ \citenamefont {Min}}]{park2022revealing}%
  \BibitemOpen
  \bibfield  {author} {\bibinfo {author} {\bibfnamefont {J.}~\bibnamefont
  {Park}}, \bibinfo {author} {\bibfnamefont {H.}~\bibnamefont {Cho}}, \bibinfo
  {author} {\bibfnamefont {S.}~\bibnamefont {Lee}}, \bibinfo {author}
  {\bibfnamefont {K.}~\bibnamefont {Lee}}, \bibinfo {author} {\bibfnamefont
  {K.}~\bibnamefont {Lee}}, \bibinfo {author} {\bibfnamefont {H.~C.}\
  \bibnamefont {Park}}, \bibinfo {author} {\bibfnamefont {J.-W.}\ \bibnamefont
  {Ryu}}, \bibinfo {author} {\bibfnamefont {N.}~\bibnamefont {Park}}, \bibinfo
  {author} {\bibfnamefont {S.}~\bibnamefont {Jeon}},\ and\ \bibinfo {author}
  {\bibfnamefont {B.}~\bibnamefont {Min}},\ }\bibfield  {title} {\bibinfo
  {title} {Revealing non-{H}ermitian band structure of photonic {F}loquet
  media},\ }\href {https://doi.org/10.1126/sciadv.abo6220} {\bibfield
  {journal} {\bibinfo  {journal} {Science Advances}\ }\textbf {\bibinfo
  {volume} {8}},\ \bibinfo {pages} {eabo6220} (\bibinfo {year}
  {2022})}\BibitemShut {NoStop}%
\bibitem [{\citenamefont {Wang}\ \emph {et~al.}(2025)\citenamefont {Wang},
  \citenamefont {Garg}, \citenamefont {Mirmoosa}, \citenamefont {Lamprianidis},
  \citenamefont {Rockstuhl},\ and\ \citenamefont
  {Asadchy}}]{wang2025expanding}%
  \BibitemOpen
  \bibfield  {author} {\bibinfo {author} {\bibfnamefont {X.}~\bibnamefont
  {Wang}}, \bibinfo {author} {\bibfnamefont {P.}~\bibnamefont {Garg}}, \bibinfo
  {author} {\bibfnamefont {M.~S.}\ \bibnamefont {Mirmoosa}}, \bibinfo {author}
  {\bibfnamefont {A.~G.}\ \bibnamefont {Lamprianidis}}, \bibinfo {author}
  {\bibfnamefont {C.}~\bibnamefont {Rockstuhl}},\ and\ \bibinfo {author}
  {\bibfnamefont {V.~S.}\ \bibnamefont {Asadchy}},\ }\bibfield  {title}
  {\bibinfo {title} {Expanding momentum bandgaps in photonic time crystals
  through resonances},\ }\href {https://doi.org/10.1038/s41566-024-01563-3}
  {\bibfield  {journal} {\bibinfo  {journal} {Nature Photonics}\ }\textbf
  {\bibinfo {volume} {19}},\ \bibinfo {pages} {149} (\bibinfo {year}
  {2025})}\BibitemShut {NoStop}%
\bibitem [{\citenamefont {Mart{\'\i}nez-Romero}\ and\ \citenamefont
  {Halevi}(2017)}]{martinez2017standing}%
  \BibitemOpen
  \bibfield  {author} {\bibinfo {author} {\bibfnamefont {J.~S.}\ \bibnamefont
  {Mart{\'\i}nez-Romero}}\ and\ \bibinfo {author} {\bibfnamefont
  {P.}~\bibnamefont {Halevi}},\ }\bibfield  {title} {\bibinfo {title} {Standing
  waves with infinite group velocity in a temporally periodic medium},\ }\href
  {https://doi.org/10.1103/PhysRevA.96.063831} {\bibfield  {journal} {\bibinfo
  {journal} {Physical Review A}\ }\textbf {\bibinfo {volume} {96}},\ \bibinfo
  {pages} {063831} (\bibinfo {year} {2017})}\BibitemShut {NoStop}%
\bibitem [{\citenamefont {Reyes-Ayona}\ and\ \citenamefont
  {Halevi}(2015)}]{reyes2015observation}%
  \BibitemOpen
  \bibfield  {author} {\bibinfo {author} {\bibfnamefont {J.~R.}\ \bibnamefont
  {Reyes-Ayona}}\ and\ \bibinfo {author} {\bibfnamefont {P.}~\bibnamefont
  {Halevi}},\ }\bibfield  {title} {\bibinfo {title} {Observation of genuine
  wave vector ($k$ or $\beta$) gap in a dynamic transmission line and temporal
  photonic crystals},\ }\href {https://doi.org/10.1063/1.4928659} {\bibfield
  {journal} {\bibinfo  {journal} {Applied Physics Letters}\ }\textbf {\bibinfo
  {volume} {107}},\ \bibinfo {pages} {074101} (\bibinfo {year}
  {2015})}\BibitemShut {NoStop}%
\bibitem [{\citenamefont {Wang}\ \emph {et~al.}(2023)\citenamefont {Wang},
  \citenamefont {Mirmoosa}, \citenamefont {Asadchy}, \citenamefont {Rockstuhl},
  \citenamefont {Fan},\ and\ \citenamefont {Tretyakov}}]{wang2023metasurface}%
  \BibitemOpen
  \bibfield  {author} {\bibinfo {author} {\bibfnamefont {X.}~\bibnamefont
  {Wang}}, \bibinfo {author} {\bibfnamefont {M.~S.}\ \bibnamefont {Mirmoosa}},
  \bibinfo {author} {\bibfnamefont {V.~S.}\ \bibnamefont {Asadchy}}, \bibinfo
  {author} {\bibfnamefont {C.}~\bibnamefont {Rockstuhl}}, \bibinfo {author}
  {\bibfnamefont {S.}~\bibnamefont {Fan}},\ and\ \bibinfo {author}
  {\bibfnamefont {S.~A.}\ \bibnamefont {Tretyakov}},\ }\bibfield  {title}
  {\bibinfo {title} {Metasurface-based realization of photonic time crystals},\
  }\href {https://doi.org/10.1126/sciadv.adg7541} {\bibfield  {journal}
  {\bibinfo  {journal} {Science Advances}\ }\textbf {\bibinfo {volume} {9}},\
  \bibinfo {pages} {eadg7541} (\bibinfo {year} {2023})}\BibitemShut {NoStop}%
\bibitem [{\citenamefont {Xiong}\ \emph {et~al.}(2025)\citenamefont {Xiong},
  \citenamefont {Zhang}, \citenamefont {Duan}, \citenamefont {Wang},
  \citenamefont {Long}, \citenamefont {Hou}, \citenamefont {Yu}, \citenamefont
  {Zou},\ and\ \citenamefont {Zhang}}]{xiong2025observation}%
  \BibitemOpen
  \bibfield  {author} {\bibinfo {author} {\bibfnamefont {J.}~\bibnamefont
  {Xiong}}, \bibinfo {author} {\bibfnamefont {X.}~\bibnamefont {Zhang}},
  \bibinfo {author} {\bibfnamefont {L.}~\bibnamefont {Duan}}, \bibinfo {author}
  {\bibfnamefont {J.}~\bibnamefont {Wang}}, \bibinfo {author} {\bibfnamefont
  {Y.}~\bibnamefont {Long}}, \bibinfo {author} {\bibfnamefont {H.}~\bibnamefont
  {Hou}}, \bibinfo {author} {\bibfnamefont {L.}~\bibnamefont {Yu}}, \bibinfo
  {author} {\bibfnamefont {L.}~\bibnamefont {Zou}},\ and\ \bibinfo {author}
  {\bibfnamefont {B.}~\bibnamefont {Zhang}},\ }\bibfield  {title} {\bibinfo
  {title} {Observation of wave amplification and temporal topological state in
  a non-synthetic photonic time crystal},\ }\href
  {https://doi.org/10.1038/s41467-025-66154-4} {\bibfield  {journal} {\bibinfo
  {journal} {Nature Communications}\ }\textbf {\bibinfo {volume} {16}},\
  \bibinfo {pages} {11182} (\bibinfo {year} {2025})}\BibitemShut {NoStop}%
\bibitem [{\citenamefont {Hou}\ \emph {et~al.}(2025)\citenamefont {Hou},
  \citenamefont {Peng}, \citenamefont {Wang}, \citenamefont {Wang},
  \citenamefont {Zhang}, \citenamefont {Wang}, \citenamefont {Hu},\ and\
  \citenamefont {Xiong}}]{hou2026experimental}%
  \BibitemOpen
  \bibfield  {author} {\bibinfo {author} {\bibfnamefont {H.}~\bibnamefont
  {Hou}}, \bibinfo {author} {\bibfnamefont {K.}~\bibnamefont {Peng}}, \bibinfo
  {author} {\bibfnamefont {Y.}~\bibnamefont {Wang}}, \bibinfo {author}
  {\bibfnamefont {J.}~\bibnamefont {Wang}}, \bibinfo {author} {\bibfnamefont
  {X.}~\bibnamefont {Zhang}}, \bibinfo {author} {\bibfnamefont
  {R.}~\bibnamefont {Wang}}, \bibinfo {author} {\bibfnamefont {H.}~\bibnamefont
  {Hu}},\ and\ \bibinfo {author} {\bibfnamefont {J.}~\bibnamefont {Xiong}},\
  }\href {https://doi.org/10.48550/arXiv.2511.14190} {\bibinfo {title}
  {Experimental realization of a full-band wave antireflection based on
  temporal taper metamaterials}} (\bibinfo {year} {2025}),\ \bibinfo {note}
  {preprint (update journal volume/pages when published)},\ \Eprint
  {https://arxiv.org/abs/2511.14190} {arXiv:2511.14190 [physics.optics]}
  \BibitemShut {NoStop}%
\bibitem [{\citenamefont {Zhu}\ \emph {et~al.}(2025)\citenamefont {Zhu},
  \citenamefont {Huang}, \citenamefont {Xu}, \citenamefont {Chen},
  \citenamefont {Meng}, \citenamefont {Zhu}, \citenamefont {Xi},\ and\
  \citenamefont {Gao}}]{zhu2025spatiotemporal}%
  \BibitemOpen
  \bibfield  {author} {\bibinfo {author} {\bibfnamefont {Z.}~\bibnamefont
  {Zhu}}, \bibinfo {author} {\bibfnamefont {B.}~\bibnamefont {Huang}}, \bibinfo
  {author} {\bibfnamefont {S.}~\bibnamefont {Xu}}, \bibinfo {author}
  {\bibfnamefont {J.}~\bibnamefont {Chen}}, \bibinfo {author} {\bibfnamefont
  {Y.}~\bibnamefont {Meng}}, \bibinfo {author} {\bibfnamefont {Z.}~\bibnamefont
  {Zhu}}, \bibinfo {author} {\bibfnamefont {X.}~\bibnamefont {Xi}},\ and\
  \bibinfo {author} {\bibfnamefont {Z.}~\bibnamefont {Gao}},\ }\href
  {https://doi.org/10.48550/arXiv.2512.16516} {\bibinfo {title} {Spatiotemporal
  topological phase transitions in photonic spacetime crystals}} (\bibinfo
  {year} {2025}),\ \bibinfo {note} {preprint},\ \Eprint
  {https://arxiv.org/abs/2512.16516} {arXiv:2512.16516 [physics.optics]}
  \BibitemShut {NoStop}%
\bibitem [{\citenamefont {Dong}\ \emph {et~al.}(2025)\citenamefont {Dong},
  \citenamefont {Chen},\ and\ \citenamefont {Yuan}}]{4lqd-z567}%
  \BibitemOpen
  \bibfield  {author} {\bibinfo {author} {\bibfnamefont {Z.}~\bibnamefont
  {Dong}}, \bibinfo {author} {\bibfnamefont {X.}~\bibnamefont {Chen}},\ and\
  \bibinfo {author} {\bibfnamefont {L.}~\bibnamefont {Yuan}},\ }\bibfield
  {title} {\bibinfo {title} {Extremely narrow band in {Moir{\'e}} photonic time
  crystal},\ }\href {https://doi.org/10.1103/4lqd-z567} {\bibfield  {journal}
  {\bibinfo  {journal} {Physical Review Letters}\ }\textbf {\bibinfo {volume}
  {135}},\ \bibinfo {pages} {033803} (\bibinfo {year} {2025})}\BibitemShut
  {NoStop}%
\bibitem [{\citenamefont {Zou}\ \emph {et~al.}(2024)\citenamefont {Zou},
  \citenamefont {Hu}, \citenamefont {Wu}, \citenamefont {Long}, \citenamefont
  {Chong}, \citenamefont {Zhang},\ and\ \citenamefont {Luo}}]{zou2024momentum}%
  \BibitemOpen
  \bibfield  {author} {\bibinfo {author} {\bibfnamefont {L.}~\bibnamefont
  {Zou}}, \bibinfo {author} {\bibfnamefont {H.}~\bibnamefont {Hu}}, \bibinfo
  {author} {\bibfnamefont {H.}~\bibnamefont {Wu}}, \bibinfo {author}
  {\bibfnamefont {Y.}~\bibnamefont {Long}}, \bibinfo {author} {\bibfnamefont
  {Y.}~\bibnamefont {Chong}}, \bibinfo {author} {\bibfnamefont
  {B.}~\bibnamefont {Zhang}},\ and\ \bibinfo {author} {\bibfnamefont
  {Y.}~\bibnamefont {Luo}},\ }\href {https://doi.org/10.48550/arXiv.2411.00215}
  {\bibinfo {title} {Momentum flatband and superluminal propagation in a
  photonic time {Moir{\'e}} superlattice}} (\bibinfo {year} {2024}),\ \bibinfo
  {note} {preprint},\ \Eprint {https://arxiv.org/abs/2411.00215}
  {arXiv:2411.00215 [physics.optics]} \BibitemShut {NoStop}%
\bibitem [{\citenamefont {Pan}\ \emph {et~al.}(2023)\citenamefont {Pan},
  \citenamefont {Cohen},\ and\ \citenamefont {Segev}}]{pan2023superluminal}%
  \BibitemOpen
  \bibfield  {author} {\bibinfo {author} {\bibfnamefont {Y.}~\bibnamefont
  {Pan}}, \bibinfo {author} {\bibfnamefont {M.-I.}\ \bibnamefont {Cohen}},\
  and\ \bibinfo {author} {\bibfnamefont {M.}~\bibnamefont {Segev}},\ }\bibfield
   {title} {\bibinfo {title} {Superluminal $k$-gap solitons in nonlinear
  photonic time crystals},\ }\href
  {https://doi.org/10.1103/PhysRevLett.130.233801} {\bibfield  {journal}
  {\bibinfo  {journal} {Physical Review Letters}\ }\textbf {\bibinfo {volume}
  {130}},\ \bibinfo {pages} {233801} (\bibinfo {year} {2023})}\BibitemShut
  {NoStop}%
\bibitem [{\citenamefont {Pendry}\ \emph {et~al.}(2021)\citenamefont {Pendry},
  \citenamefont {Galiffi},\ and\ \citenamefont {Huidobro}}]{pendry2021gain}%
  \BibitemOpen
  \bibfield  {author} {\bibinfo {author} {\bibfnamefont {J.~B.}\ \bibnamefont
  {Pendry}}, \bibinfo {author} {\bibfnamefont {E.}~\bibnamefont {Galiffi}},\
  and\ \bibinfo {author} {\bibfnamefont {P.~A.}\ \bibnamefont {Huidobro}},\
  }\bibfield  {title} {\bibinfo {title} {Gain in time-dependent media---a new
  mechanism},\ }\href {https://doi.org/10.1364/JOSAB.427682} {\bibfield
  {journal} {\bibinfo  {journal} {Journal of the Optical Society of America B}\
  }\textbf {\bibinfo {volume} {38}},\ \bibinfo {pages} {3360} (\bibinfo {year}
  {2021})}\BibitemShut {NoStop}%
\bibitem [{SM()}]{SM}%
  \BibitemOpen
  \href@noop {} {}\bibinfo {note} {See Supplemental Material for the
  calculation of the bound on the energy-transport velocity, the cycle-averaged
  flux and energy density, the wavepacket centroid and its decomposition, and
  the extension to dispersive and weakly lossy PTCs.}\BibitemShut {Stop}%
\bibitem [{\citenamefont {Johnson}\ \emph {et~al.}(2002)\citenamefont
  {Johnson}, \citenamefont {Ibanescu}, \citenamefont {Skorobogatiy},
  \citenamefont {Weisberg}, \citenamefont {Joannopoulos},\ and\ \citenamefont
  {Fink}}]{JohnsonIb02}%
  \BibitemOpen
  \bibfield  {author} {\bibinfo {author} {\bibfnamefont {S.~G.}\ \bibnamefont
  {Johnson}}, \bibinfo {author} {\bibfnamefont {M.}~\bibnamefont {Ibanescu}},
  \bibinfo {author} {\bibfnamefont {M.~A.}\ \bibnamefont {Skorobogatiy}},
  \bibinfo {author} {\bibfnamefont {O.}~\bibnamefont {Weisberg}}, \bibinfo
  {author} {\bibfnamefont {J.~D.}\ \bibnamefont {Joannopoulos}},\ and\ \bibinfo
  {author} {\bibfnamefont {Y.}~\bibnamefont {Fink}},\ }\bibfield  {title}
  {\bibinfo {title} {Perturbation theory for {Maxwell}'s equations with
  shifting material boundaries},\ }\href
  {https://doi.org/10.1103/PhysRevE.65.066611} {\bibfield  {journal} {\bibinfo
  {journal} {Physical Review E}\ }\textbf {\bibinfo {volume} {65}},\ \bibinfo
  {pages} {066611} (\bibinfo {year} {2002})}\BibitemShut {NoStop}%
\bibitem [{\citenamefont {Kato}(1995)}]{Kato1995Perturbation}%
  \BibitemOpen
  \bibfield  {author} {\bibinfo {author} {\bibfnamefont {T.}~\bibnamefont
  {Kato}},\ }\href {https://doi.org/10.1007/978-3-642-66282-9} {\emph {\bibinfo
  {title} {Perturbation Theory for Linear Operators}}},\ \bibinfo {edition}
  {2nd}\ ed.,\ Classics in Mathematics\ (\bibinfo  {publisher} {Springer},\
  \bibinfo {address} {Berlin, Heidelberg},\ \bibinfo {year} {1995})\BibitemShut
  {NoStop}%
\end{thebibliography}

\begin{thebibliography}{5}%
\makeatletter
\providecommand \@ifxundefined [1]{%
 \@ifx{#1\undefined}
}%
\providecommand \@ifnum [1]{%
 \ifnum #1\expandafter \@firstoftwo
 \else \expandafter \@secondoftwo
 \fi
}%
\providecommand \@ifx [1]{%
 \ifx #1\expandafter \@firstoftwo
 \else \expandafter \@secondoftwo
 \fi
}%
\providecommand \natexlab [1]{#1}%
\providecommand \enquote  [1]{``#1''}%
\providecommand \bibnamefont  [1]{#1}%
\providecommand \bibfnamefont [1]{#1}%
\providecommand \citenamefont [1]{#1}%
\providecommand \href@noop [0]{\@secondoftwo}%
\providecommand \href [0]{\begingroup \@sanitize@url \@href}%
\providecommand \@href[1]{\@@startlink{#1}\@@href}%
\providecommand \@@href[1]{\endgroup#1\@@endlink}%
\providecommand \@sanitize@url [0]{\catcode `\\12\catcode `\$12\catcode
  `\&12\catcode `\#12\catcode `\^12\catcode `\_12\catcode `\%12\relax}%
\providecommand \@@startlink[1]{}%
\providecommand \@@endlink[0]{}%
\providecommand \url  [0]{\begingroup\@sanitize@url \@url }%
\providecommand \@url [1]{\endgroup\@href {#1}{\urlprefix }}%
\providecommand \urlprefix  [0]{URL }%
\providecommand \Eprint [0]{\href }%
\providecommand \doibase [0]{https://doi.org/}%
\providecommand \selectlanguage [0]{\@gobble}%
\providecommand \bibinfo  [0]{\@secondoftwo}%
\providecommand \bibfield  [0]{\@secondoftwo}%
\providecommand \translation [1]{[#1]}%
\providecommand \BibitemOpen [0]{}%
\providecommand \bibitemStop [0]{}%
\providecommand \bibitemNoStop [0]{.\EOS\space}%
\providecommand \EOS [0]{\spacefactor3000\relax}%
\providecommand \BibitemShut  [1]{\csname bibitem#1\endcsname}%
\let\auto@bib@innerbib\@empty
\bibitem [{\citenamefont {Kato}(1995)}]{SM-Kato1995Perturbation}%
  \BibitemOpen
  \bibfield  {author} {\bibinfo {author} {\bibfnamefont {T.}~\bibnamefont
  {Kato}},\ }\href {https://doi.org/10.1007/978-3-642-66282-9} {\emph {\bibinfo
  {title} {Perturbation Theory for Linear Operators}}},\ \bibinfo {edition}
  {2nd}\ ed.,\ Classics in Mathematics\ (\bibinfo  {publisher} {Springer},\
  \bibinfo {address} {Berlin, Heidelberg},\ \bibinfo {year} {1995})\BibitemShut
  {NoStop}%
\bibitem [{\citenamefont {Sundaram}\ and\ \citenamefont
  {Niu}(1999)}]{SM-SundaramNiu1999WavepacketDynamics}%
  \BibitemOpen
  \bibfield  {author} {\bibinfo {author} {\bibfnamefont {G.}~\bibnamefont
  {Sundaram}}\ and\ \bibinfo {author} {\bibfnamefont {Q.}~\bibnamefont {Niu}},\
  }\bibfield  {title} {\bibinfo {title} {Wave-packet dynamics in slowly
  perturbed crystals: Gradient corrections and {B}erry-phase effects},\ }\href
  {https://doi.org/10.1103/PhysRevB.59.14915} {\bibfield  {journal} {\bibinfo
  {journal} {Physical Review B}\ }\textbf {\bibinfo {volume} {59}},\ \bibinfo
  {pages} {14915} (\bibinfo {year} {1999})}\BibitemShut {NoStop}%
\bibitem [{\citenamefont {Park}\ \emph {et~al.}(2025)\citenamefont {Park},
  \citenamefont {Lee}, \citenamefont {Zhang}, \citenamefont {Park},
  \citenamefont {Ryu}, \citenamefont {Cho}, \citenamefont {Lee}, \citenamefont
  {Zhang}, \citenamefont {Park}, \citenamefont {Jeon}, \citenamefont {Shin},
  \citenamefont {Chan},\ and\ \citenamefont {Min}}]{SM-park2025spontaneous}%
  \BibitemOpen
  \bibfield  {author} {\bibinfo {author} {\bibfnamefont {J.}~\bibnamefont
  {Park}}, \bibinfo {author} {\bibfnamefont {K.}~\bibnamefont {Lee}}, \bibinfo
  {author} {\bibfnamefont {R.-Y.}\ \bibnamefont {Zhang}}, \bibinfo {author}
  {\bibfnamefont {H.-C.}\ \bibnamefont {Park}}, \bibinfo {author}
  {\bibfnamefont {J.-W.}\ \bibnamefont {Ryu}}, \bibinfo {author} {\bibfnamefont
  {G.~Y.}\ \bibnamefont {Cho}}, \bibinfo {author} {\bibfnamefont {M.~Y.}\
  \bibnamefont {Lee}}, \bibinfo {author} {\bibfnamefont {Z.}~\bibnamefont
  {Zhang}}, \bibinfo {author} {\bibfnamefont {N.}~\bibnamefont {Park}},
  \bibinfo {author} {\bibfnamefont {W.}~\bibnamefont {Jeon}}, \bibinfo {author}
  {\bibfnamefont {J.}~\bibnamefont {Shin}}, \bibinfo {author} {\bibfnamefont
  {C.~T.}\ \bibnamefont {Chan}},\ and\ \bibinfo {author} {\bibfnamefont
  {B.}~\bibnamefont {Min}},\ }\bibfield  {title} {\bibinfo {title} {Spontaneous
  emission decay and excitation in photonic time crystals},\ }\href
  {https://doi.org/10.1103/5v2w-yg7v} {\bibfield  {journal} {\bibinfo
  {journal} {Physical Review Letters}\ }\textbf {\bibinfo {volume} {135}},\
  \bibinfo {pages} {133801} (\bibinfo {year} {2025})}\BibitemShut {NoStop}%
\bibitem [{\citenamefont {Landau}\ \emph {et~al.}(1984)\citenamefont {Landau},
  \citenamefont {Lifshitz},\ and\ \citenamefont
  {Pitaevskii}}]{SM-LandauLifshitzECM}%
  \BibitemOpen
  \bibfield  {author} {\bibinfo {author} {\bibfnamefont {L.~D.}\ \bibnamefont
  {Landau}}, \bibinfo {author} {\bibfnamefont {E.~M.}\ \bibnamefont
  {Lifshitz}},\ and\ \bibinfo {author} {\bibfnamefont {L.~P.}\ \bibnamefont
  {Pitaevskii}},\ }\href@noop {} {\emph {\bibinfo {title} {Electrodynamics of
  Continuous Media}}},\ \bibinfo {edition} {2nd}\ ed.,\ \bibinfo {series}
  {Course of Theoretical Physics}, Vol.~\bibinfo {volume} {8}\ (\bibinfo
  {publisher} {Pergamon Press},\ \bibinfo {address} {Oxford, UK},\ \bibinfo
  {year} {1984})\BibitemShut {NoStop}%
\bibitem [{\citenamefont
  {Jackson}(1999)}]{SM-Jackson1998ClassicalElectrodynamics}%
  \BibitemOpen
  \bibfield  {author} {\bibinfo {author} {\bibfnamefont {J.~D.}\ \bibnamefont
  {Jackson}},\ }\href@noop {} {\emph {\bibinfo {title} {Classical
  Electrodynamics}}},\ \bibinfo {edition} {3rd}\ ed.\ (\bibinfo  {publisher}
  {Wiley},\ \bibinfo {address} {New York, NY},\ \bibinfo {year} {1999})\
  \bibinfo {note} {often cited as 1998; copyright/print year commonly listed as
  1999}\BibitemShut {NoStop}%
\end{thebibliography}
\end{document}